%% file: ms.tex
\documentclass{emulateapj}
\usepackage{times}
\usepackage{lscape}

\def\la{\mathrel{\mathpalette\fun <}}
\def\ga{\mathrel{\mathpalette\fun >}}

\def\fun#1#2{\lower0.837ex\vbox{\baselineskip0ex\lineskip0.209ex
  \ialign{$\mathsurround=0ex#1\hfil##\hfil$\crcr#2\crcr\sim\crcr}}}

\def\sles{\lower2pt\hbox{$\buildrel {\scriptstyle <}
   \over {\scriptstyle\sim}$}}

\def\sgreat{\lower2pt\hbox{$\buildrel {\scriptstyle >}
   \over {\scriptstyle\sim}$}}

\def\la{\mathrel{\mathpalette\fun <}}
\def\ga{\mathrel{\mathpalette\fun >}}

\begin{document}

\title{Correlations of Prompt and Afterglow 
    Emission in {\it Swift} Long and Short Gamma Ray Bursts}
\shortauthors{GEHRELS et al.}
\author{
N.~Gehrels\altaffilmark{1},
S.~D.~Barthelmy\altaffilmark{1},
D.~N.~Burrows\altaffilmark{2},
J.~K.~Cannizzo\altaffilmark{1,3},
G.~Chincarini\altaffilmark{4,5},
E.~Fenimore\altaffilmark{6},
C.~Kouveliotou\altaffilmark{7},
P.~O'Brien\altaffilmark{8},
D.~M.~Palmer\altaffilmark{6},
J.~Racusin\altaffilmark{2},
P.~W.~A.~Roming\altaffilmark{2},
T.~Sakamoto\altaffilmark{1,3},
J.~Tueller\altaffilmark{1},
R.~A.~M.~J.~Wijers\altaffilmark{9},
B.~Zhang\altaffilmark{10}
}
\altaffiltext{1}{NASA-Goddard Space Flight Center, Greenbelt, MD 20771, 
                 neil.gehrels@nasa.gov}
\altaffiltext{2}{Department of Astronomy \& Astrophysics,
                 Pennsylvania State University, State College, PA 16802}
\altaffiltext{3}{CRESST/Joint Center for Astrophysics,
                 University of Maryland, Baltimore County, Baltimore, MD 21250}
\altaffiltext{4}{INAF-Osservatorio Astronomico di Brera, 1-23807 Merate, Italy}
\altaffiltext{5}{Universita degli studi di Milano Bicocca, 1-20126, Milano, Italy}
\altaffiltext{6}{Los Alamos National Laboratory, P.O. Box 1663, Los Alamos, NM, 87545}
\altaffiltext{7}{NASA-Marshall Space Flight Center, NSSTC, VP-62, 
                 320 Sparkman Drive, Huntsville, AL 35805}
\altaffiltext{8}{Department of Physics \& Astronomy, University of Leicester, Leicester, LE1 7RH, UK}
\altaffiltext{9}{Faculty of Science, Astronomical Institute ``Anton Pannekoek'', 
              University of Amsterdam, Kruislaan 403, 1098 SJ Amsterdam, The Netherlands}
\altaffiltext{10}{Department of Physics and Astronomy, University of 
                  Nevada Las Vegas, Las Vegas, NV 89154}


\begin{abstract}
Correlation studies of prompt and afterglow emissions
from gamma-ray bursts (GRBs) between different spectral
bands has been difficult to do in the past because 
few bursts had comprehensive and intercomparable afterglow 
measurements.  In this paper\footnote{to appear in The Astrophysical Journal, Dec 20, 2008, v. 689, no. 2}
  we present a large and 
uniform data set for correlation analysis based on bursts 
detected by the {\it Swift} mission.  For the first time, short
and long bursts can be analyzed and compared.  It is found
for both classes that the optical, X-ray and gamma-ray 
emissions are linearly correlated, but with a large spread 
about the correlation line; stronger bursts tend to have 
brighter afterglows, and bursts with brighter X-ray afterglow 
tend to have brighter optical afterglow.  Short bursts are, 
on average, weaker in both prompt and afterglow emissions.  
No short bursts are seen with extremely low optical 
to X-ray ratio as occurs for ``dark'' long bursts.  
Although statistics are still poor for short bursts, there is 
no evidence yet for a subgroup of short bursts with high 
extinction as there is for long bursts.  Long bursts are 
detected in the dark category at the same fraction as for 
pre-{\it Swift} bursts.  Interesting cases are discovered of 
long bursts that are detected in the optical, and yet have 
low enough optical to X-ray ratio to be classified as dark. 
 For the prompt emission, short and long bursts have different 
average tracks on flux {\it vs} fluence plots.  In {\it Swift}, GRB 
detections tend to be fluence limited for short bursts and 
flux limited for long events.
\end{abstract}


\section{Introduction}


One of the longest enduring Gamma Ray Burst (GRB) classification 
schemes is based on their distributions in duration and spectral 
hardness.  Both quantities seem to cluster into two separate 
classes with the longer events (those above $\sim2$ s; 
Kouveliotou et al. 1993) being predominantly softer while the 
shorter ones are harder. The mechanism for the origin of the 
GRB explosions (the central engine) appears to be quite 
different for the two types. 
Long bursts are ascribed to the 
core collapse to a black hole of a massive, young, rapidly 
rotating star in the ``collapsar'' model (Woosley 1993; 
MacFadyen \& Woosley 1999; Woosley \& Bloom 2006) which is 
supported by observations such as the coincidence of SNe with 
well-observed nearby GRBs (Galama et al. 1998; 
Bloom et al. 1999; Staneck et al. 2003; Hjorth et al. 2003;
Pian et al. 2006). 
 The prevalent model for short bursts  has  them        caused 
by the coalescence of a binary pair of compact old stars 
(Lattimer \& Schramm 1974; Paczy\'nski 1986; Eichler et al. 1989; 
Mochkovitch et al. 1993; Rosswog, Ramirez-Ruiz, \& Davies 2003;
Oechslin, Janka, \& Marek  2007) which is supported by recent observations 
of progenitor sites with low star formation activity 
(Gehrels et al. 2005; Bloom et al. 2006; Fox et al. 2005; 
Villasenor et al. 2005, Hjorth et al. 2005; 
Barthelmy et al. 2005b; Berger et al. 2005).  In both scenarios, 
a highly-relativistic collimated outflow of particles and 
radiation       occurs producing prompt gamma-ray emission 
from shock accelerated electrons, which evolves into a 
long-lasting afterglow from shock interactions with 
the circumburst medium (e.g., M\'esz\'aros \& Rees 1997).  
For short bursts there are also models for the afterglow in 
which a radioactive wind causes emission in the first 
day or so (Li \& Paczy\'nski 1998, Kulkarni 2005).

Correlation studies of prompt and afterglow emission 
are crucial for understanding their production mechanisms 
and environmental effects.  For example, Jakobsson 
et al. (2004) developed a criterion for determining which GRBs 
are ``dark'' bursts, by comparing the relative intensity of 
their X-ray and optical afterglows to find what fraction of 
bursts have high column densities. 
Stratta et al. (2004) studied the X-ray and optical absorption
properties of 13 GRBs studied by {\it BeppoSax}.
 Roming et al. (2006) and 
Fynbo et al. (2007) expanded on previous work to include (long) bursts 
from the {\it Swift} satellite.  A more detailed work on dark bursts 
using a broad-band spectral analysis is given by 
Rol et al. (2005, 2007).  Zhang et al. (2007) present a study 
comparing radiative efficiencies for short and long bursts as 
derived from a correlation analysis.  Using {\it Swift} short bursts,
 Berger (2007) compared their X-ray afterglow to their 
gamma-ray prompt emission, and found that 20\% have anomalously 
low X-ray to gamma ray ratios indicating very low density 
burst sites, possibly in globular clusters, for that 
subpopulation (see also Berger et al 2007).
   Other correlation studies
   have been undertaken by
   Salmonson \& Galama (2002),
   Firmani et al. (2006),
   Nava et al. (2006),
   Butler (2007), and
   Nysewander, Fruchter, \& Pe'er (2008).
An early study of X-ray afterglow properties at $t=11$ hr
was carried out by Piran et al. (2001).

In this study we perform correlation studies using the
 extensive data set from {\it Swift}.  Sections 2 and 3 cover
 observations and results, respectively, while in Section 4, 
we discuss the implications of the results and in Section 5 
the conclusions and future prospects.

\section{Observations}

\subsection{{\it Swift} Studies}

The {\it Swift} mission (Gehrels et al. 2004) has so far 
provided {\it uniform observations 
of prompt and afterglow emission}
for hundreds of GRBs.  This sample is an order of magnitude 
larger than the one previously available with e.g., the 
{\it BeppoSAX} satellite
(de Pasquale et al. 2006).\footnote{see also website by J. Greiner: 
 http://www.mpe.mpg.de/$\sim$jcg/grbgen.html}
Further, {\it Swift} X-ray observations 
covering time-scales from 1 minute to several days 
after the burst are provided for the first time 
for   most every GRBs.  After three years of operations, 
our data set has now reached a critical size where 
statistically meaningful correlations can be studied.

We present here three correlation studies:  
(1) X-ray {\it vs} optical afterglow, 
(2) gamma-ray prompt 
{\it vs} X-ray afterglow, and 
(3) prompt gamma-ray peak 
flux {\it vs} fluence. 
               All the data used in this study are listed 
in Tables $1-4$ except that gamma-ray data are 
listed only for 
those bursts with, at least, an X-ray afterglow.  The full list 
of fluences and fluxes for the 193 bursts used for study (3) are 
directly from the Sakamoto et al. (2008) tables.  
We include all {\it Swift} bursts from January 2005 
through July 2007 for studies (1) and (2) and through 
February 2007 for study (3).
   We adopt $T90 = 2$ s for the dividing line
   between long and short GRBs,
   except for ones with soft extended emission.
   In those cases
   the duration of the initial hard pulse was required to be 
   $<2$ s, and only that emission was used in the analysis
   (GRB 050724, 051227, 061006, 061210, and 070714B).
   Including the extended emission
   in the fluence would increase it by a factor
   $\la 2$         and would not significantly
   change the correlations.

For the X-ray {\it vs} optical afterglow study, we use the 
methods developed by Jakobsson et al. (2004) in their 
comparison of X-ray and optical afterglow fluxes for 
pre-{\it Swift} bursts.  In order to compare to the 
Jakobsson et al. results, we use the same definition 
of quantities:  the X-ray flux density at 3 keV, the optical flux in the 
$R-$band, and sampling time at 11 hr after the burst. 
 The {\it Swift} X-ray lightcurves have been found typically 
to have complex shapes (Nousek et al. 2005; Zhang et al. 2005) 
often including a poorly understood ``plateau phase''; the use of 
flux at 11 hr in most cases avoids sampling during 
the plateau phase and gives a measure of the true burst afterglow.

\subsection{X-ray Fluxes}

The X-ray fluxes are from measurements 
of the {\it Swift} X-Ray Telescope 
(XRT; Burrows et al. 2005).  Our primary data product 
for the XRT flux is the integral flux between 
0.3 and 10 keV corrected for absorption at low 
energies (unabsorbed flux).  This is converted 
to the flux density at 3 keV using the measured 
spectral index. 
Given an integral $0.3-10$ keV X-ray flux $[I_x]=$ erg cm$^{-2}$ s$^{-1}$
    and a 
$0.3-10$ keV X-ray photon index $n$, 
the flux density at 3 keV, in $\mu$Jy, is given by
\begin{equation}
f_X(3 \ {\rm keV}) = 4.13\times 10^{11} {I_x (2-n) E_0^{1-n} \over{\left( { E_2^{2-n} - E_1^{2-n}  } \right) }},
\end{equation}
where $E_0=  3$ keV,
      $E_1=0.3$ keV, and
      $E_2=10$ keV.
%
%
%
%
%
 The integral fluxes, photon spectral 
indices and flux densities are listed 
in Tables $1-3$.
    A 10\% systematic uncertainty was 
added in quadrature to the measured error to account 
for uncertainties in the shape and variability of the lightcurves.

The integral flux calculation was carried out as follows
(see J. Racusin et al. 2008, in preparation,
for a more detailed discussion of the method).
 Level 1 data products were downloaded from 
the NASA/GSFC {\it Swift} Data Center (SDC) and 
processed using XRTDAS software (v2.0.1).  The 
{\tt xrtpipeline} task was used to generate 
level 2 cleaned event files.  Only events with 
Windowed Timing (WT) mode grades $0-2$ and 
Photon Counting (PC) mode grades $0-12$ and energies 
between $0.3-10.0$ keV were used in subsequent 
temporal and spectral analysis.

The XRT light curves were created by extracting 
the counts in a circular region around 
the GRB afterglow with a variable source 
radius designed to optimize the $S/N$ depending 
on the count rate.  They were background subtracted, 
pile-up corrected where applicable, 
exposure map corrected, and corrected for
the fraction of the PSF excluded by the extraction region.  
The number of counts per bin is variable and dependent 
on the count rate.  Time intervals of significant 
flaring were removed from the light curves and 
they were fit to power-laws, broken power-laws, and 
multiply broken power-law.  
Using these temporal fits, we 
interpolated the count rate at 11 hr.

Spectra for the power-law segments of the light curves 
were extracted individually to limit contamination 
by potential spectral variability.  The segment used 
for the counts to flux conversion was that
 at  11 hr.  The spectra were created by extracting 
the counts in a 20 pixel radius extraction region and 
a 40 pixel radius background region.  
The Ancillary Response Files were made using the {\tt xrtmkarf}
task and grouped with 20 counts per bin using the 
{\tt grppha} task.  The spectra were fit in XSPEC to absorbed 
power-laws and used to measure the $0.3-10$ keV flux and 
count rate which was applied to the interpolated count 
rate to convert into flux units.

\subsection{Optical and Gamma Ray Fluxes}

The optical fluxes are from measurements by ground-based 
telescopes and from the {\it Swift} UV Optical Telescope 
(UVOT; Roming et al. 2005).  An extensive literature search was done 
to find the best optical data for each burst.  Bursts were included 
in the study if measurements were available within a factor of 2 of 
11 hr (i.e., at $>5.5$ hr or $<22$ hr).  The value at 11 hr was 
estimated by interpolations and extrapolations when measurements 
were not available exactly at 11 hr. The correction applied to the $R$ 
data for $t_{\rm obs}\ne$ 11 hr was 
  $\Delta m_R = -2.5\log_{10}(t_{\rm obs}/11. \ {\rm hr})$.
 The one exception to the 
factor of 2 criterion was GRB 070508 with measurements to only 4 hr, 
which was included because it appears to be an interesting 
dark burst candidate.  A few bursts are listed with optical 
flux upper limits at the bottom of Table 2.  This is not an 
exhaustive list of optical limits, but only those with low optical 
to X-ray ratio limits.  A 10\% systematic uncertainty was added in 
quadrature to the measured error to account for uncertainties in 
the shape and variability of the lightcurves.

Galactic extinction was taken into account
using the study of
 Schlegel, Finkbeiner, \& Davis (1998).\footnote{http://nedwww.ipac.caltech.edu/forms/calculator.html} 
For the precise sky map positions
  we utilize  the XRT localizations.
For each data source reference, 
    a determination had to be made as to whether the
galactic correction had already been made. (For the GCN entries, it
  was always assumed the correction had not been made.)
 For most of the GRBs, the $R$ band correction is small (a few tenths
of a magnitude). The exceptions from Table 1 are 
       050724 ($\Delta m_R = 1.64$)
               and
       061006 ($\Delta m_R = 0.85$);
                 the exceptions from Table 2 are 
       050713B ($\Delta m_R = 1.249$) and
       070704  ($\Delta m_R = 5.014$).
  Corrections this large are highly uncertain due to 
the patchiness in extinction in the Galactic plane.

   Our sample contains three  GRBs with redshift values
large enough ($z\simeq4$)  so that Lyman blanketing may affect the $R-$band fluxes.
 For these bursts $-$
                    050730, $z=3.97$; 
                    060206, $z=4.05$; and 
                    060210, $z=3.91$ $-$  
    the expected redshifting of the Lyman series $(1+z)\simeq5$.
 For  Ly$\alpha$,  $1215.7  \AA \ \rightarrow \ \sim 6080 \AA$
   and  for
      Ly$\infty$,  $911.3 \AA \ \rightarrow \ \sim  4560 \AA$.
Thus the effect of the redshifted  absorption is to
impact the blue edge of the  $R-$band filter
  $\lambda_R \approx 6600 \pm  800 \AA$.
          The $R-$band fluxes for the three highest $z$ bursts
(indicated by circles in Fig. 1)
scatter about the mean $R-$band flux line, however, 
     rather than being concentrated
at low $F_R$ values as would have been expected had blanketing been an  
issue.
Although for $z=4$ the Ly$\alpha$ feature will be shifted redward of the 
$\sim 6000 \AA$ (skew-symmetric)
   peak of the $R$ filter, the centroid and FWHM of the filter
still predict the bulk of the filter response
  to lie redward of most of the  Lyman series,
  which would be most dominant for  $z=4$ at  $\lambda < 6000 \AA$
   (e.g.,
for the Cousins [Bessell 90] $R_C$ filter,
    $\lambda_{\rm eff}=6588 \AA$ and $\Delta\lambda_{\rm FWHM} =1568 \AA$ $-$
Fukugita, Shimasaku, \& Ichikawa 1995, see their Table 9).
The absorptive effect would likely warrant a corrective multiplicative factor $\la1.5$, 
  which is small
given the $\sim 5$ decade spread in $F_R$ for Figure 1.  Therefore we do not  
attempt to correct $F_R$ for redshifted 
         Lyman absorption for these three high-z bursts.

\begin{figure}
\centering
\epsscale{1.205}
\plotone{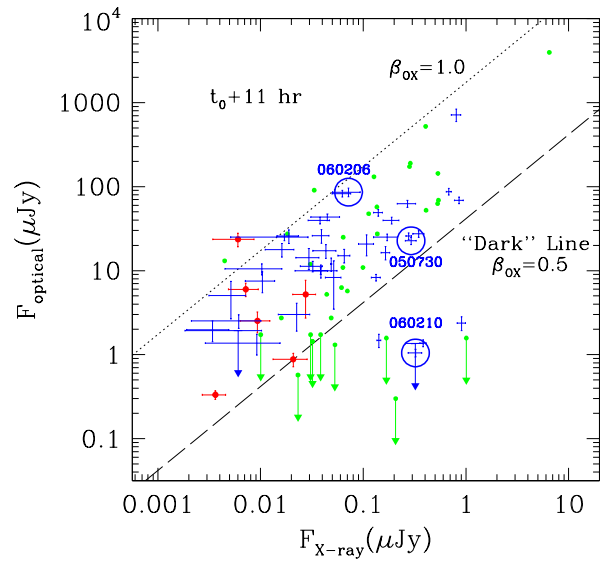}
\vskip -2cm
\figcaption{
The
 optical afterglow 
{\it vs} 
  X-ray afterglow 
flux densities of {\it Swift}
short (shown in red) and long (shown in blue) GRBs
 at 11 hr after the burst.  The three circled bursts are those
for which $z>3.9$.
Also plotted are the pre-{\it Swift}
 GRBs (shown in green) taken from Jakobsson et al (2004).
For the Jakobsson et al subsample with upper limits,
we only plot those bursts for which the limiting magnitude
is fainter than $m_R=23$ (i.e., $\sim2$ $\mu$Jy).
  The XRT X-ray flux densities 
are at 3 keV and the optical flux densities are in the $R-$band 
(see Table 1 and 2).  Also shown is the ``dark'' burst 
    separation     line $\beta_{OX} = 0.5$ (Jakobsson et al 2004),
 and a line indicating  $\beta_{OX} = 1.0$.
\label{fig1}}
\smallskip
\end{figure}

The gamma-ray fluences and peak fluxes are in the $15-150$ keV band 
and are from the {\it Swift} Burst Alert Telescope 
(BAT; Barthelmy et al. 2005a) as compiled in the BAT GRB catalog 
(Sakamoto et al. 2008).  For the gamma-ray flux needed in 
study (3), we use 1 s binning as quoted by Sakamoto et al. (2008).

\subsection{Correlation Analysis}

For each study, we have performed fits to the two-parameter correlation 
data using the Spearman rank test (Spearman 1904; Press et al. 1986) and derived 
the correlation coefficient, $r$, to determine the degree and significance 
of the correlation.  Upper limits were not included in the fits.  
In the Spearman rank test, the probability 
  of a null correlation, $P_{\rm null}$, is given by
\begin{equation}
P_{\rm null} = {\rm erfc} [r (N/2)^{1/2}]
\end{equation}
where $N$ is the number of data points.  The significance of 
the correlation is $P_{\rm cor} = 1 - P_{\rm null}$.  
The fraction of the observed spread of the data that 
can be explained by the correlation is given by $r^2$.  
The fit parameters and correlation $r$ values are listed in Table 4.
 Equation (2)  only applies in the limit  of $N$ large ($\ga10-20$).
  For $N\simeq1-10$, the concept of applying a significance criterion
to a correlation study begins to lose its
  meaning.\footnote{This can be seen 
in the limit $N\rightarrow 2$ 
           where one considers two data points $(x_1,y_1)$ and
 $(x_2,y_2)$.  In this example
       $r\equiv 1$,
     so the statement ``$r=1$'' 
 carries no information and has no significance.}
 Therefore, although 
for completeness
      we list $r$ and $P_{\rm null}$ values for
cases with small $N$, we stress that 
they are only indicative of trends in those cases. 
%

\section{Results}

\subsection{X-ray and Optical Afterglow Correlations}

Figure 1 shows the {\it Swift} X-ray afterglow average 
flux density 
at 3 keV as a function of the 
$R-$band optical flux density, both 
converted to $\mu$Jy at 11 hr after the burst, for short and long bursts. 
The pre-{\it Swift} data points are taken from 
Jakobsson et al. (2004) and are shown as filled green points.  
Also shown is the solid line of constant X-ray to optical spectral 
index that they propose separates the true ``dark'' bursts 
from the rest.  As listed in Table 4, the Spearman rank test for the 
two GRB populations in Figure 1 gives a null probability 
of $\sim0.01$ or a  99\% correlation probability between the optical 
and X-ray flux densities of the long GRBs, 
          and only $\sim30$\% for the short population.  

The long {\it Swift} GRBs fall in the same general region of 
the plot as the pre-{\it Swift} ones.  As with the pre-{\it Swift} bursts, 
several {\it Swift} long bursts (detections and upper limits) 
also fall below the Jakobsson et al. dark line.  The brightest 
short GRBs fall     in the midst of the long GRB points, 
but in the region toward lower flux densities. 
  To 
date there are no short bursts that fall below the dark burst line; 
those with low optical flux densities or upper limits tend to also have 
weak X-ray flux densities that place them above the line.

\subsection{Gamma Ray Prompt and X-ray Afterglow Correlations}

\begin{figure}
\centering
\epsscale{1.205}
\plotone{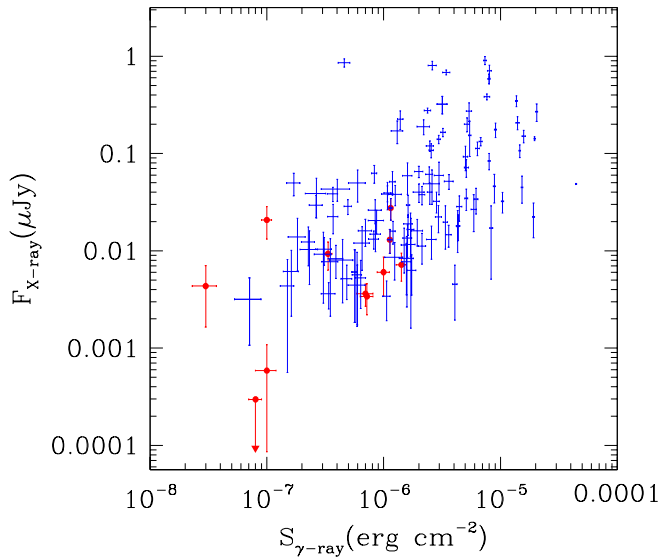}
\vskip -2cm
\figcaption{
 The X-ray afterglow flux density {\it vs} gamma-ray prompt fluence of {\it Swift}
 short (shown in red) and long (shown in blue) GRBs at 11 hr after the burst.
 The XRT X-ray flux densities are at 3 keV and the BAT
gamma-ray fluences are between 15 and 150 keV (Sakamoto et al. 2008).
    The XRT and BAT
data are given in Table 1, 2 and 3.
\label{fig2}}
\smallskip
\end{figure}

We show in Figure 2 the average X-ray 
   afterglow flux density {\it vs} the gamma-ray 
fluence of the prompt emission for long and short {\it Swift} GRBs.  
We find a highly significant correlation (99.9999996\% probability) for 
the long GRBs, albeit with a wide spread in the data.  The correlation 
of the short bursts is less significant (69\% probability) mostly due 
to the smaller number of points.  There is an overlap between 
the brightest short bursts and the faintest long GRBs.  
The weakest short bursts are fainter than the weakest long bursts.

\subsection{Prompt Gamma Ray Fluence and Peak Flux Correlations}

\begin{figure}
\centering
\epsscale{1.205}
\plotone{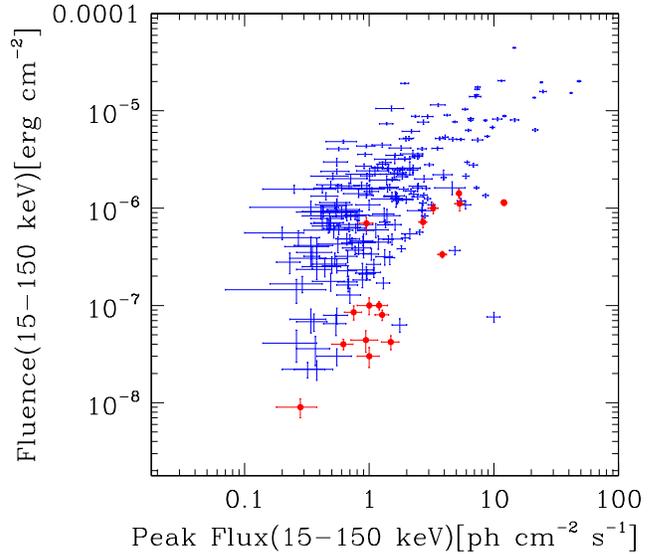}
\vskip -2cm
\figcaption{
The prompt gamma-ray fluence {\it vs} peak flux measured by 
BAT in the 15 to 150 keV band for all bursts through 
February 2007 (Sakamoto et al. 2008). 
 Short bursts are shown by red symbols
and long bursts by blue.
\label{fig3}}
\smallskip
\end{figure}

Figure 3 shows the prompt emission fluence as a function of peak 
flux for GRBs detected by BAT.  We see a linear correlation for 
both short and long bursts with a significant spread in the 
correlation. The correlation probability is virtually 100\% 
(null probability $= 2\times 10^{-29}$) for long bursts and 99.9998\% 
for short bursts.  The best fit lines are distinctly different 
for short and long bursts, with the long burst having a higher 
fluence on average for a given flux level than 
short bursts as expected from duration alone.

\section{Discussion}

\subsection{Correlations \& Short/Long Distributions}

We show in this work that correlations exist between 
prompt and afterglow fluxes of GRBs and between different 
wavelength bands in the afterglow.  The highest significance 
correlation is between the prompt emission gamma-ray fluence 
and the X-ray afterglow flux at a significance level 
of 99.9999996\% for long bursts and 69\% for short bursts.  
The correlation between the optical afterglow and 
X-ray afterglow fluxes is less significant at 99\% 
significance for long bursts and only $\sim30$\% for 
short bursts (for a small sample, however).  

It is important to note that there is a 
wide spread in the data for all of the correlations.  
The correlations are real and significant, but the fraction 
of the observed variations due to the correlations 
between the above parameters accounts for only a portion 
of the data spread.  The correlation can only be used 
to predict a flux level to within approximately an 
order of magnitude.  The fraction of the variation due 
to the correlations is given by the square of the 
correlation parameter, $r$, which, as shown in Table 4, 
varies from a few percent to 50\%.  The rest of the data 
spread is due to other factors such as correlations with 
additional unknown parameters.  An example of an additional 
parameter is extinction in the optical afterglow.

Short bursts are weaker on average than long bursts in 
afterglow fluxes.  There is overlap with the dimmer 
long bursts, but the short bursts extend to lower intensities 
than seen for long bursts.  The average X-ray flux density at 3 keV 
at 11 hr for the short bursts is $<F_x({\rm short})> = 9.6\times10^{-3}\mu$Jy, 
which is more than an order of magnitude less than 
the average for long bursts of    $<F_x({\rm long})> = 0.10\mu$Jy.

 The X-ray to gamma-ray correlation in Figure 2 has a positive correlation with a slope of roughly unity. This
suggests that brighter bursts have more kinetic energy  in the afterglow phase to power the afterglow. This is a
manifestation of similar radiative efficiency among different bursts and between long and short GRBs. Such a point was
made by Zhang et al. (2007) based on an analysis of a smaller sample of early {\it Swift} GRBs.

Except for the bursts below the ``dark'' line, most bursts in Figure 1 are confined between lines with 
$\beta_{OX} =$
0.5 and 1.0. This is consistent with a general interpretation that 
the optical and X-ray emission belong to the same
spectral component with an index close to 0.75.
Within the standard model for emission via synchrotron radiation,
 for slow cooling 
which is generally relevant at 
  $t = 11$ hr, one
expects $\beta_{OX} \sim (p-1)/2$ for $\nu_m < \nu_O < \nu_X < \nu_c$, 
which has a typical value of 0.75 for
    electron distribution power law $p = 2.5$. 
   (An equivalent statement is that 
     for this spectrum, the predicted ratio  
    $F_R/F_X \approx 350$ yields a
    line intermediate between 
    $\beta_{OX}=0.5$ and $1.0$ in 
    Fig. 1.) 
   This
suggests that on average, the cooling frequency is above or 
not much below the X-ray band at 11 hr.

\subsection{Dark GRBs}

Another comparison of short and long GRBs relates to dark bursts.  
Jakobsson et al. (2004, see also De Pasquale et al. 2003)
       used the simple criterion to define dark bursts 
as those with extremely low optical to X-ray afterglow ratio, falling 
below the line of optical to X-ray spectral index, $\beta_{\rm OX}$, equal to 0.5. 
 It may seem counterintuitive that there can be dark bursts 
with optical detections and bursts not detected in the optical 
that are not ``dark'', but the important criterion is how 
optically faint the burst is relative to its X-ray flux.  
For the pre-{\it Swift} sample there were 5 bursts with 
upper limits below the dark-burst line
(restricting the Jakobsson et al. sample
  to include only those with 
upper limits fainter than
$m_R=23$, or $\sim2\mu$Jy),
compared to 24 bursts with 
actual measurements (not upper limits) above the line, giving a fraction 
of $\sim17$\% in the dark category.  For {\it Swift} there are 2 bursts 
with upper limits (GRB 050713B and  061222A) and 
 3 cases with measurements (GRB 060210, 070419B and 070508) below 
the line compared with 34 long  bursts above the line for 
a fraction of $\sim17$\% in the dark category,     the same as 
the pre-{\it Swift} sample.  The conclusion is that {\it Swift}
 is sampling the same source environments as previous   instruments.

The discovery of 3 cases of dark bursts with optical detections is 
particularly interesting.  One possible concern with this finding is 
that {\it Swift} X-ray afterglows are contaminated in many bursts 
by emission components not from the external shocks, e.g. X-ray flares.
  In such instances, the Jakobsson et al. (2004) approach 
to define dark bursts is no longer relevant since it assumes that 
the X-ray and optical emission is from the same emission component, 
but separated by a cooling break.  However, the X-ray lightcurves 
for the {\it Swift} dark bursts are smooth around 11 hr
(and beyond the end of the X-ray plateau),
  with 
no significant contamination from other components.
 These are real ``dark'' bursts from both an observational 
and physics perspective.




Correlation analyses between optical and X-ray can help answer 
the question of whether these two afterglow
components originate from the same physical processes. 
 It is assumed in the Jakobsson et al (2004) study
that both X-ray and optical emission arise from the external forward
shock. Multiwavelength observations in the {\it Swift} era reveal 
puzzling chromatic features of
afterglow breaks (e.g., Panaitescu et al. 2006; Liang et al. 2007, 2008) 
that are not consistent
with the simplest forward shock model. Models invoking 
non-forward-shock origin of X-ray
afterglows have been discussed in the literature 
(e.g., Genet et al. 2007; Uhm \& Beloborodov
2007; Ghisellini et al. 2007; Shao \& Dai 2007; Panaitescu 2008). 
On the other hand, analyses
suggest that the X-ray data are generally consistent with 
the temporal index and spectral
index relations (e.g., Zhang \& M\'esz\'aros 2004) predicted by 
the forward shock models, although not in
 every case. (Liang
et al. 2007,
Willingale, et al. 2007).
   The optical/X-ray data of some bursts 
(e.g., Grupe et al. 2007; Mangano et
al. 2007) are      consistent with the same forward shock model. 
   Regardless of the exact process, 
 the analysis presented in
this paper shows that generally optical/X-ray afterglow 
fluxes are correlated, which suggests that they are due to
the same emission process. The few cases well below the 
correlation line are found to be dark due to extinction
in the host galaxy.

For the first time we can search for dark short bursts.  
No short bursts are seen that fall below the dark-burst line.  
It is hard to find dark GRBs using this criterion since X-ray 
afterglow fluxes are also low for the short bursts.  However, there are 
some short bursts with bright X-ray afterglow, and, to date, none of 
those is seen to be highly deficient in optical afterglow.  Statistics 
are still small with only 5 optical detections, but if the observed 
trend continues we will be able to conclude that short bursts do not 
occur in regions with extremely high extinction as occurs for some 
long bursts.

%

 We are beginning to have optical detections of bursts 
below the dark burst line.
In one of the three dark bursts with
detections (GRB 060210), the burst
is found to have high extinction associated with its host galaxy,
explaining the low optical flux (Curran et al. 2007b). 
 By modeling the differences between $\beta_{\rm opt}$, $\beta_X$, and $\beta_{OX}$, 
 and taking into
 account the Lyman$-\alpha$ absorption ($z=3.91$), 
the authors find the $R-$band source extinction could be $3.9\pm0.7$ mag
 ($\nu_c > \nu_O$) or $6.7\pm0.6$ mag   ($\nu_c < \nu_O$).
 This is an important development
in our understanding of dark GRBs.  
 (For two of the three dark bursts with
detections - 070419B and 070508 -
there has not yet been sufficiently detailed follow-up work
on the putative hosts for constraints to be placed on the host extinction.)
Assuming that the dark bursts can be largely explained by extinction, then the
optical - X-ray correlations, ignoring the dark bursts, would hold true.
    We note that new studies are being done to examine
  dark burst definitions (van der Horst et al. 2008).

%
%
%

\subsection{Prompt Fluence and Flux Comparisons}

The comparison of fluences and peak fluxes in the prompt emission as 
shown in Figure 3 is a different kind of study than in the other two 
above.  In this case, the strong observed correlation and high degree 
of separation of short and long bursts is expected; brighter bursts 
with higher peak fluxes naturally have higher fluences and short bursts 
tend to have lower fluence for a given flux by the very fact of their 
short duration.  Within the short and long classes, the spread in 
fluence that is seen for a given peak flux is due to the diversity 
of durations and spectral indices.  Bursts with longer duration and 
hard spectra have higher fluences for the same peak flux.

It is interesting to note in Figure 3 that the short bursts tend to be 
fluence limited in the BAT, while long bursts tend to be peak 
flux limited.  This is due to the way BAT operates.  A valid GRB trigger 
requires a statistically significant excess in both the rate 
and image domains (Fenimore et al. 2004).  The ability to form an 
image depends on the number of photons collected on various 
trigger timescales, which is related to the burst fluence.  Even for 
relatively high peak-fluxes, short bursts can have low fluence  values 
and be limited in the number of photons available for the image trigger.
  On the other hand, long bursts tend to have higher fluences for a 
given peak-flux and become rate limited before the image limit 
is reached.  BAT also has a pure-image mode for triggering where 
very long duration GRBs and other transients are found by comparing 
sky images instead of having a rate trigger.  The lowest long-burst 
point in Figure 3 at a peak flux of $\sim0.1$ was such an image-mode 
trigger for the very long ($T90 = 35$ min) and weak GRB 060218.  
A caveat on the above discussion is that the BAT trigger algorithm 
is complex with $\sim500$ different trigger criteria evaluated.  
There are many different thresholds and limits coming into play 
for short and long burst triggering, with some mix of flux and 
fluence limits for both types.

This study was based on a 1 s binning for the gamma-ray fluxes.  
We have also investigated the effect of using a smaller 
bin size of 64 ms.  Smaller bins pick out larger peak flux values 
when there is short time structure or when the burst has a duration 
shorter than the bin size.  The effect of the smaller bin size 
is to shift the short bursts to the right (higher peak flux) 
relative to the long bursts by about a factor of 5.  
The larger bin size that we use allows for
    better statistics and is more reliable for long bursts.  
In either case, the short bursts tend to cluster toward lower 
fluences than long bursts.

\section{Future Prospects}

The combined prompt and afterglow data set for {\it Swift} GRBs is the 
largest available to date.  We have chosen a criterion on the 
afterglow measurements for inclusion in this study of being a 
solid measurement 11 hr after the burst.  Even with this 
stringent definition, there are more than 100 long bursts with 
X-ray afterglow data.  The optical detections at 11 hr are 
less numerous with about 40 good measurements, but still enough 
statistics for conclusions to be reached.

The short burst correlation  studies are possible now and key 
results are beginning to emerge.  The {\it Swift} data base is 
growing quickly with time.  In its expected lifetime of 
$\sim10$ yr, the mission should provide a sample of 
$>40$ short and $>400$ long GRBs with good afterglow and 
prompt observations. 
That sized data set will allow more detailed correlations 
studies to investigate the interesting trends found in
the current analysis.

\acknowledgements

The authors thank the following people for useful discussions: 
A. Fruchter, J. Fynbo, P. Jakobsson, A. Loeb, M. Nysewander, and E. Rol.
We also acknowledge our anonymous referee who
provided excellent suggestions for improving the paper.


\def\mnras{MNRAS}
\def\apj{ApJ}
\def\apjs{ApJS}
\def\apjl{ApJL}
\def\aj{AJ}
\def\araa{ARA\&A}
\def\aap{A\&A}

\vfil\eject
\pagestyle{empty}
\setlength{\voffset}{20mm}
\LongTables
\clearpage
\begin{landscape}
\voffset=-.235truein
\input{tab1.tex}
\clearpage
\end{landscape}

\vfil\eject
\pagestyle{empty}
\setlength{\voffset}{20mm}
\LongTables
\clearpage
\begin{landscape}
\voffset=-.235truein
\input{tab2.tex}
\clearpage
\end{landscape}

\vfil\eject
\clearpage
\pagestyle{empty}
\setlength{\voffset}{-15mm}
\LongTables
\clearpage
\input{tab3.tex}
\clearpage

\vfil\eject
\clearpage
\pagestyle{empty}
\setlength{\voffset}{20mm}
\input{tab4.tex}
\clearpage

\end{document}

%% file: tab1.tex
\begin{deluxetable}{lcccccccccrcccrr}
\tabletypesize{\tiny}
\tablewidth{0pt}
 \tablecaption{{\it Swift} Short GRBs with X-ray or Optical Data at 11 Hr.}
\tablehead{
\colhead{(1)} & 
\colhead{(2)} & 
\colhead{(3)} & 
\colhead{(4)} & 
\colhead{(5)} & 
\colhead{(6)}  & 
\colhead{(7)}  & 
\colhead{(8)}  & 
\colhead{(9)}  & 
\colhead{(10)} & 
\colhead{(11)} & 
\colhead{(12)} & 
\colhead{(13)} & 
\colhead{(14)} & 
\colhead{(15)} & 
\colhead{(16)} \\
\colhead{GRB} &
\colhead{$\gamma$-ray\tablenotemark{a}} & 
\colhead{$\gamma$-ray\tablenotemark{a}} & 
\colhead{X-ray}  & 
\colhead{X-ray} &  
\colhead{X-ray} &  
\colhead{X-ray} &  
\colhead{X-ray} &  
\colhead{X-ray\tablenotemark{b}} &  
\colhead{X-ray\tablenotemark{b}} &  
\colhead{$m_R$} &  
\colhead{$m_R$} &  
\colhead{Time\tablenotemark{c}} &    
\colhead{$R$ Flux\tablenotemark{d}}& 
\colhead{$R$ Flux\tablenotemark{d}} & 
\colhead{$R$} \\    
\colhead{ }      &  
\colhead{Fluence} & 
\colhead{Fluence} & 
\colhead{Integral}& 
\colhead{Integral}& 
\colhead{Spectrum}& 
\colhead{Spectrum}& 
\colhead{Spectrum}& 
\colhead{Flux}  &   
\colhead{Flux}  &   
\colhead{ }       & 
\colhead{Error}   & 
\colhead{$R-$band}& 
\colhead{Density} & 
\colhead{Density} & 
\colhead{Ref.} \\   
\colhead{ }    &  
\colhead{ }     & 
\colhead{Error} & 
\colhead{Flux}&   
\colhead{Flux} &  
\colhead{Photon}& 
\colhead{Photon}& 
\colhead{Photon}& 
\colhead{Density}& 
\colhead{Density}&
\colhead{}      &
\colhead{    }  &
\colhead{data}  &
\colhead{@11 hr}&
\colhead{Error} &
\colhead{} \\    
\colhead{ } & 
\colhead{ } & 
\colhead{ } & 
\colhead{$0.3-10$ keV} & 
\colhead{Error} & 
\colhead{Index} & 
\colhead{Index} & 
\colhead{Index} & 
\colhead{@3 keV} & 
\colhead{Error}  & 
\colhead{}       & 
\colhead{}       & 
\colhead{or code}& 
\colhead{      } & 
\colhead{     }  & 
\colhead{} \\      
\colhead{ } & 
\colhead{ } & 
\colhead{ } & 
\colhead{@11 hr} & 
\colhead{} & 
\colhead{} & 
\colhead{Lower} & 
\colhead{Upper} & 
\colhead{@11 hr} & 
\colhead{}  & 
\colhead{}  & 
\colhead{ } & 
\colhead{ } & 
\colhead{}   &
\colhead{}  & 
\colhead{} \\ 
\colhead{ } & 
\colhead{} & 
\colhead{} & 
\colhead{} & 
\colhead{} & 
\colhead{} & 
\colhead{Error} & 
\colhead{Error} & 
\colhead{}  & 
\colhead{}  & 
\colhead{ } & 
\colhead{ } & 
\colhead{}  & 
\colhead{ } & 
\colhead{ } & 
\colhead{} \\ 
\colhead{ } & 
\colhead{$10^{-7}$} & 
\colhead{$10^{-7}$} & 
\colhead{$10^{-13}$} & 
\colhead{$10^{-13}$} & 
\colhead{     } & 
\colhead{     } & 
\colhead{     } & 
\colhead{$10^{-3}$} & 
\colhead{$10^{-3}$} & 
\colhead{     }  & 
\colhead{      }  & 
\colhead{  }       & 
\colhead{       }  & 
\colhead{       }  & 
\colhead{ } \\       
\colhead{ } & 
\colhead{erg cm$^{-2}$} & 
\colhead{erg cm$^{-2}$} & 
\colhead{erg cm$^{-2}$ s$^{-1}$} & 
\colhead{erg cm$^{-2}$ s$^{-1}$} & 
\colhead{     } & 
\colhead{     } & 
\colhead{     } & 
\colhead{$\mu$Jy} & 
\colhead{$\mu$Jy} & 
\colhead{     }  & 
\colhead{      }  & 
\colhead{hr}       & 
\colhead{$\mu$Jy}  & 
\colhead{$\mu$Jy}  & 
\colhead{ } \\       
}
\startdata
050509B & 0.09 & 0.02 & $-$  & $-$  & $-$  & $-$  & $-$  & $-$  & $-$  & $>21.9$ & UL  & 11.  & $<5.0$  & UL  & \ref{tv1}\\
050724  & 9.98 &  1.2 & 1.59 & 0.51 & 2.06 & 0.52 & 0.79 & 6.03 &  2.6 & 20.3    & 0.2 & Ff   & 23.6    & 4.4 & \ref{tv2}\\
050813  & 0.44 & 0.11 & $-$  & $-$  & $-$  & $-$  & $-$  & $-$  & $-$  & $>22.5$ & UL  & 13.  & $<3.1$  & UL  & \ref{tv3}\\
051220  & 0.85 & 0.14 & $-$  & $-$  & $-$  & $-$  & $-$  & $-$  & $-$  & $-$     & $-$ & $-$  & $-$     & $-$ &          \\
051221A & 11.5 & 0.35 & 7.52 & 1.4  & 2.12 & 0.17 & 0.19 & 27.5 & 6.9  & 21.9    & 0.5 & If   &  5.22   & 2.5 & \ref{tv4}\\
051227  & 6.99 & 1.1  & 0.864& 0.18 & 1.86 & 0.21 & 0.23 & 3.63 & 0.93 & 24.9    & 0.12& 11.45&  0.333 & 0.04 & \ref{ber}\\
060313  & 11.3 & 0.45 & 4    & 0.66 & 2.27 & 0.2  & 0.23 & 13   & 3.6  & $>20.6$ & UL  & 8.4  & $<17.2$ & UL &\ref{twx1a}\\
060502B & 0.4  & 0.05 & $-$  & $-$  & $-$  & $-$  & $-$  & $-$  & $-$  & $>23.2$ & UL  & 16.8 & $<1.54$ & UL  & \ref{tv7}\\
060801  & 0.8  &0.1 &$<0.139$& UL   & 2.69 & 0.69 & 1.1& $<0.297$ & UL & $-$     & $-$ & $-$  & $-$     & $-$ &          \\
061006  & 14.2 & 1.4 & 1.66  & 0.43 & 1.78 & 0.28 & 0.44 & 7.17 & 2.3  & 21.8    & 0.2 & 14.6 & 6.0     & 1.1 & \ref{mal}\\
061201  & 3.34 & 0.27& 2.07  & 0.59 & 1.61 & 0.27 & 0.55 & 9.32 & 3    & 22.7    & 0.3 & 8.38 & 2.52    & 0.7 & \ref{tv9}\\
061217  & 0.42 & 0.07 & 2.17 & 3.1  & $-$  & $-$  & $-$  & $-$  & $-$  & $-$     & $-$ & $-$  & $-$     & $-$ &           \\
070724A & 0.3  & 0.07 & 1.18 & 0.53 & 2.1  & 0.46 & 0.57 & 4.34 & 2.7  & $-$     & $-$ & $-$  & $-$     & $-$ &           \\
070729  & 1    & 0.2  & 0.16 & 0.091& 2.11 & 0.77 & 1.5  & 0.586& 0.5  & $-$     & $-$ & $-$  & $-$     & $-$ &           \\
070809  & 1    & 0.1  & 4.55 & 1.5  & 1.4  & 0.35 & 0.5  & 20.8 & 7.6  & 23.8    & 0.2 & 11.21& 0.879   & 0.16& \ref{tv12}\\
\enddata
\tablenotetext{a}{BAT prompt fluence in $15-150$ keV band. Data from Sakamoto et al. (2008).}
\tablenotetext{b}{XRT flux density at 3 keV at 11 hr after the burst trigger. Error includes 10\% systematic uncertainty.}
\tablenotetext{c}{Hrs after burst trigger of optical data, or code for optical data. If a number,
       it is the time after the burst (typically listed for GCN only data).
       If letters, the first letter is for F = full light curve, I = interpolated between
          measured values on either side of 11 hr, and E = extrapolated from measured data. 
 The lower case letters indicate if the data in the referenced papers was in 
magnitude or Jansky units (mmf = first two references in magnitude units, 
third reference in Jansky units)}
\tablenotetext{d} {Optical data in $R-$band at 11 hr after the burst trigger. 
 $R-$band flux density estimated from typical burst spectra if data taken in other bands.  
Error includes 10\% systematic uncertainty.}

 
 \newcounter{refno} 
 \setcounter{refno}{0} 
\tablerefs{
 \refstepcounter{refno} \label{tv1}     [\arabic{refno}] Misra \& Pandey (2005)
 \refstepcounter{refno} \label{tv2}     [\arabic{refno}] Malesani et al. (2007)
 \refstepcounter{refno} \label{tv3}     [\arabic{refno}]  Bikmaev et al. (2005)
 \refstepcounter{refno} \label{tv4}     [\arabic{refno}] Soderberg et al. (2006)
 \refstepcounter{refno} \label{ber}     [\arabic{refno}] Berger et al. (2007)
 \refstepcounter{refno} \label{twx1a}   [\arabic{refno}] Schmidt \& Bayliss (2006)
 \refstepcounter{refno} \label{tv7}     [\arabic{refno}] Price et al. (2006)
 \refstepcounter{refno} \label{mal}     [\arabic{refno}] Malesani et al. (2006)
 \refstepcounter{refno} \label{tv9}     [\arabic{refno}] D'Avanzo et al. (2006)
 \refstepcounter{refno} \label{tv12}    [\arabic{refno}] Perley, Thoene, \& Bloom (2007)
}
\end{deluxetable}

%% file: tab2.tex
\begin{deluxetable}{lcccccccccrcccrr}
\tabletypesize{\tiny}
\tablewidth{0pt}
 \tablecaption{{\it Swift} Long GRBs with X-ray Detections and Optical Detections
          or Low Upper Limits at 11 Hr.}
\tablehead{
\colhead{(1)} & 
\colhead{(2)} & 
\colhead{(3)} & 
\colhead{(4)} & 
\colhead{(5)} & 
\colhead{(6)}  & 
\colhead{(7)}  & 
\colhead{(8)}  & 
\colhead{(9)}  & 
\colhead{(10)} & 
\colhead{(11)} & 
\colhead{(12)} & 
\colhead{(13)} & 
\colhead{(14)} &
\colhead{(15)} &
\colhead{(16)} \\
\colhead{GRB} &
\colhead{$\gamma$-ray\tablenotemark{a}} & 
\colhead{$\gamma$-ray\tablenotemark{a}} & 
\colhead{X-ray}  & 
\colhead{X-ray} &  
\colhead{X-ray} &  
\colhead{X-ray} &  
\colhead{X-ray} &  
\colhead{X-ray\tablenotemark{b}} & 
\colhead{X-ray\tablenotemark{b}} & 
\colhead{$m_R$} &  
\colhead{$m_R$} &  
\colhead{Time\tablenotemark{c}} &    
\colhead{$R$ Flux\tablenotemark{d}}& 
\colhead{$R$ Flux\tablenotemark{d}}& 
\colhead{$R$} \\    
\colhead{ }      &  
\colhead{Fluence} & 
\colhead{Fluence} & 
\colhead{Integral}& 
\colhead{Integral}& 
\colhead{Spectrum}& 
\colhead{Spectrum}& 
\colhead{Spectrum}& 
\colhead{Flux}  &   
\colhead{Flux}  &  
\colhead{ }      & 
\colhead{Error}  & 
\colhead{$R-$band}&
\colhead{Density}& 
\colhead{Density}& 
\colhead{Ref.} \\  
\colhead{ }  &  
\colhead{ } & 
\colhead{Error} & 
\colhead{Flux}& 
\colhead{Flux} & 
\colhead{Photon}& 
\colhead{Photon}& 
\colhead{Photon}& 
\colhead{Density}& 
\colhead{Density}&
\colhead{}      & 
\colhead{    }  & 
\colhead{data}  & 
\colhead{@11 hr}& 
\colhead{Error} & 
\colhead{} \\     
\colhead{ } & 
\colhead{ } & 
\colhead{ } & 
\colhead{$0.3-10$ keV} & 
\colhead{Error} & 
\colhead{Index} & 
\colhead{Index} & 
\colhead{Index} & 
\colhead{@3 keV}& 
\colhead{Error} & 
\colhead{}      & 
\colhead{ }     &
\colhead{or code}&
\colhead{      } &
\colhead{     }  &
\colhead{} \\     
\colhead{ } & 
\colhead{ } & 
\colhead{ } & 
\colhead{@11 hr}&
\colhead{} & 
\colhead{} & 
\colhead{Lower}&
\colhead{Upper}&
\colhead{@11 hr}&
\colhead{}    & 
\colhead{}     &
\colhead{ }    &
\colhead{ }   & 
\colhead{}     &
\colhead{}    & 
\colhead{}  \\  
\colhead{ } & 
\colhead{} & 
\colhead{} & 
\colhead{} & 
\colhead{}   & 
\colhead{ }  & 
\colhead{Error}& 
\colhead{Error}& 
\colhead{}     & 
\colhead{}     & 
\colhead{ }    & 
\colhead{ }    & 
\colhead{}     & 
\colhead{  }   & 
\colhead{  }   & 
\colhead{ } \\   
\colhead{ } & 
\colhead{$10^{-7}$} & 
\colhead{$10^{-7}$} & 
\colhead{$10^{-13}$} & 
\colhead{$10^{-13}$} & 
\colhead{     } & 
\colhead{     } & 
\colhead{     } & 
\colhead{$10^{-3}$} & 
\colhead{$10^{-3}$} & 
\colhead{     }  & 
\colhead{      }  & 
\colhead{  }       & 
\colhead{       }  & 
\colhead{       }  & 
\colhead{ } \\       
\colhead{ } & 
\colhead{erg cm$^{-2}$} & 
\colhead{erg cm$^{-2}$} & 
\colhead{erg cm$^{-2}$ s$^{-1}$} & 
\colhead{erg cm$^{-2}$ s$^{-1}$} & 
\colhead{     } & 
\colhead{     } & 
\colhead{     } & 
\colhead{$\mu$Jy} & 
\colhead{$\mu$Jy} & 
\colhead{     }  & 
\colhead{      }  & 
\colhead{hr}       & 
\colhead{$\mu$Jy}  & 
\colhead{$\mu$Jy}  & 
\colhead{ } \\       
}
\startdata
{\bf with OPT} &  &       &       &      &  & &     &      &      &       &      &        &      &      &   \\
@11 hr     &      &       &       &      &  & &     &      &      &       &      &        &      &      &   \\
           &      &       &       &      &  & &     &      &      &       &      &        &      &      &   \\
050315 & 32.2 & 1.5 & 41.9 & 3.2 & 2 & 0.068 & 0.071 & 165 & 16 & 20.7 & 0.2 & 11.6 & 16.4 & 3 & 
\ref{tw1}\\
050318 & 10.8 & 0.77 & 9.61 & 1.4 & 1.93 & 0.13 & 0.14 & 39.2 & 7.3 & 20.2 & 0.2 & Ef & 26 & 4.8 & 
\ref{tw2}\\
050319 & 13.1 & 1.5 & 45 & 9.6 & 2.05 & 0.12 & 0.13 & 171 & 45 & 20.2 & 0.1 & Fmmmf & 25 & 2.3 & 
\ref{tw3a},\ref{tw3b},\ref{tw3c},\ref{tw3d}\\
050416A & 3.67 & 0.37 & 10.4 & 1.3 & 2.1 & 0.12 & 0.14 & 38.5 & 6.5 & 21.2 & 0.1 & Fff & 9.94 & 0.92 & 
\ref{tw4a},\ref{tw4b}\\
050525A & 153 & 2.2 & 14.3 & 2.7 & 2.32 & 0.2 & 0.26 & 44.8 & 14 & 19.6 & 0.1 & Ffmf & 43.4 & 4 & 
\ref{tw5a},\ref{tw5b},\ref{tw2}\\
050721 & 36.2 & 3.2 & 12.7 & 1.6 & 1.93 & 0.15 & 0.2 & 51.7 & 8.9 & 21.4 & 0.6 & If & 8.27 & 4.8 & 
\ref{tw6}\\
050730 & 23.8 & 1.5 & 62.8 & 3.3 & 1.72 & 0.051 & 0.052 & 277 & 17 & 20.2 & 0.1 & Fmff & 25.8 & 2.4 & 
\ref{tw7a},\ref{tw7b},\ref{tw2}\\
050801 & 3.1 & 0.48 & 2.47 & 0.66 & 1.85 & 0.18 & 0.28 & 10.4 & 3.3 & 21.5 & 0.3 & Ef & 7.54 & 2.1 & 
\ref{tw8}\\
050802 & 20 & 1.6 & 15.4 & 1.7 & 1.84 & 0.087 & 0.094 & 65.2 & 8.7 & 20.7 & 0.2 & Ff & 15.1 & 2.8 & 
\ref{tw2}\\
050820A & 34.4 & 2.4 & 176 & 7.5 & 2.02 & 0.048 & 0.049 & 681 & 40 & 18.8 & 0.1 & Ffff & 87.2 & 8 & 
\ref{tw10},\ref{tw7b},\ref{tw2}\\
050824 & 2.66 & 0.52 & 9.88 & 3.6 & 2.01 & 0.22 & 0.24 & 38.7 & 17 & 21.2 & 0.2 & Ff & 9.94 & 1.8 & 
\ref{tw11}\\
050908 & 4.83 & 0.51 & 1.24 & 0.38 & 1.88 & 0.28 & 0.45 & 5.15 & 2 & 21.9 & 0.5 & If & 5.06 & 2.4 & 
\ref{tw2}\\
050922C & 16.2 & 0.54 & 8.27 & 1.5 & 2.15 & 0.18 & 0.19 & 29.5 & 7.5 & 20.8 & 0.3 & 7 & 14.3 & 4 & 
\ref{tw13}\\
051109A & 22 & 2.7 & 48.5 & 6.4 & 2.02 & 0.13 & 0.14 & 189 & 33 & 19.7 & 0.1 & Fff & 39.6 & 3.7 & 
\ref{tw14},\ref{tw2}\\
060108 & 3.69 & 0.37 & 6 & 1.5 & 2.08 & 0.25 & 0.29 & 22.5 & 7.6 & 22.5 & 0.4 & If & 3 & 1.1 & 
\ref{tw16}\\
060124 & 4.61 & 0.53 & 223 & 16 & 2.04 & 0.078 & 0.081 & 856 & 83 & 19.1 & 0.1 & Fmf & 68.8 & 6.3 & 
\ref{tw17},\ref{tw2}\\
060206 & 8.31 & 0.42 & 18.6 & 2.4 & 2.23 & 0.14 & 0.16 & 62.6 & 13 & 18.9 & 0.1 & Ffffff & 82.7 & 7.6 & 
\ref{tw18a},\ref{tw18b},\ref{tw18c},            \ref{tw2}\\
060210 & 76.6 & 4.1 & 106 & 5.3 & 2.13 & 0.057 & 0.058 & 383 & 28 & 23.4 & 0.1 & Fmf & 1.37 & 0.13 & 
\ref{tw19},\ref{tw2}\\
060418 & 83.3 & 2.5 & 5.12 & 1.9 & 2.24 & 0.59 & 0.96 & 17.1 & 12 & 20.2 & 0.1 & Ff & 25 & 2.3 & 
\ref{tw7b}\\
060512 & 2.32 & 0.4 & 2.98 & 1.1 & 2.19 & 0.35 & 0.53 & 10.3 & 5.8 & 21.1 & 0.16 & 6.8 & 10.5 & 1.6 & 
\ref{tw22}\\
060526 & 12.6 & 1.6 & 8.69 & 1.7 & 1.74 & 0.17 & 0.24 & 38 & 9 & 19.7 & 0.1 & Fmff & 39.6 & 3.7 & 
\ref{tw23},            \ref{tw2}\\
060604 & 4.02 & 1.1 & 11.5 & 2.1 & 2.07 & 0.17 & 0.19 & 43.2 & 11 & 20.6 & 0.2 & 16.5 & 17.2 & 3.2 & 
\ref{tw24}\\
060605 & 6.97 & 0.9 & 4.36 & 1 & 2.1 & 0.19 & 0.27 & 16.1 & 5 & 20.6 & 0.2 & Ff & 17.8 & 3.3 & 
\ref{tw2}\\
060607A & 25.5 & 1.1 & 23.8 & 3.4 & 1.59 & 0.13 & 0.14 & 108 & 18 & 20.4 & 0.3 & Ef & 20.8 & 5.8 & 
\ref{tw26}\\
060614 & 204 & 3.6 & 70.3 & 11 & 2.04 & 0.15 & 0.16 & 269 & 55 & 19.2 & 0.1 & Ffmf & 62.7 & 5.8 & 
\ref{tw27a},\ref{tw27b},\ref{tw2}\\
060714 & 28.3 & 1.7 & 9.18 & 1.6 & 2.18 & 0.17 & 0.18 & 32.2 & 7.8 & 21.1 & 0.15 & 8.7 & 11.3 & 1.6 & 
\ref{tw28}\\
060729 & 26.1 & 2.1 & 218 & 16 & 2.1 & 0.076 & 0.078 & 805 & 84 & 16.6 & 0.18 & 20 & 716 & 120 & 
\ref{tw29}\\
060904B & 16.2 & 1.4 & 5.57 & 0.87 & 2.22 & 0.17 & 0.18 & 18.9 & 4.5 & 20.2 & 0.2 & 15.7 & 25.9 & 4.8 & 
\ref{tw30}\\
061007 & 444 & 5.6 & 11.3 & 0.11 & 1.78 & 0.014 & 0.014 & 48.7 & 0.63 & 21 & 0.2 & Im & 12 & 2.2 & 
\ref{tw31}\\
061021 & 29.6 & 1 & 37.5 & 2.7 & 2.08 & 0.079 & 0.083 & 140 & 14 & 19.5 & 0.1 & 16.5 & 49 & 4.5 & 
\ref{tw32}\\
061110A & 10.6 & 0.76 & 0.845 & 0.31 & 1.95 & 0.32 & 0.41 & 3.41 & 1.5 & 23 & 0.3 & 8 & 1.98 & 0.55 & 
\ref{tw33}\\
061121 & 137 & 2 & 82 & 8.4 & 1.84 & 0.1 & 0.11 & 347 & 44 & 20.1 & 0.1 & Fff & 27.4 & 2.5 & 
\ref{tw34},\ref{tw2}\\
061126 & 67.7 & 2.2 & 33.2 & 2.4 & 1.95 & 0.096 & 0.1 & 133 & 13 & 21.4 & 0.1 & Fm/f & 8.27 & 0.76 & 
\ref{tw35}\\
070224 & 3.05 & 0.51 & 2.49 & 1.1 & 2.1 & 0.67 & 0.91 & 9.2 & 6.3 & 23.4 & 0.3 & 7.2 & 1.37 & 0.38 & 
\ref{tw36}\\
070419B & 73.6 & 2 & 204 & 16 & 1.68 & 0.092 & 0.096 & 906 & 85 & 22.8 & 0.2 & 7.3 & 2.38 & 0.44 & 
\ref{tw37}\\
070508 & 196 & 2.7 & 32.1 & 1.2 & 1.7 & 0.054 & 0.057 & 142 & 6.5 & 23.3 & 0.2 & 4.1 & 1.48 & 0.27 & 
\ref{tw38}\\
070518 & 1.62 & 0.24 & 1.87 & 0.7 & 2.27 & 0.5 & 0.68 & 6.12 & 4 & 22.7 & 0.2 & 9 & 2.52 & 0.47 & 
\ref{tw39}\\
              &  &       &       &      &      &   & &    &      &       &      &        &      &      &   \\
              &  &       &       &      &      &   & &    &      &       &      &        &      &      &   \\
              &  &       &       &      &      &   & &    &      &       &      &        &      &      &   \\
              &  &       &       &      &      &   & &    &      &       &      &        &      &      &   \\
              &  &       &       &      &      &   & &    &      &       &      &        &      &      &   \\
              &  &       &       &      &      &   & &    &      &       &      &        &      &      &   \\
{\bf OPTICAL} &  &       &       &      &      &  & &      &      &       &      &        &      &      &   \\
{\bf LIMIT} &  &       &       &      &      &    & &   &      &       &      &        &      &      &   \\
@11 hr     &      &       &       &      &      & & &      &      &       &      &        &       &      &   \\
           &      &       &       &      &      & & &      &      &       &      &        &       &      &   \\
  050713B & 31.8 & 3.2  & 77.5  & 12    & 1.89 & 0.17 & 0.19 & 321 & 65   & $>24.6$ & UL& 6.   & $<2.72$ & UL &\ref{twx1}\\
  061004  & 5.66 & 0.31 & 2.06  & 0.67  & 2.4  & 0.53 & 0.67 & 6.04 & 4.2 & $>24.5$ & UL& 12.  & $<5.86$ & UL &\ref{twx2}\\
  061222A & 79.9 & 1.6  & 158   & 13    & 2.09 & 0.095& 0.099& 587 & 68   & $>26.1$ & UL& 16.  & $<1.69$ & UL &\ref{twx3}\\
  070721A & 0.71 & 0.18 & 1.43  & 0.45  & 2.65 & 0.39 & 0.39 & 3.17 & 2.1 & $>26.6$ & UL& 18.8 & $<1.33$ & UL &\ref{twx6}\\
\enddata
\tablenotetext{a}{BAT prompt fluence in $15-150$ keV band. Data from Sakamoto et al. (2008).}
\tablenotetext{b}{XRT flux density at 3 keV at 11 hr after the burst trigger. Error includes 10\% systematic uncertainty.}
\tablenotetext{c}{Hrs after burst trigger of optical data, or code for optical data. If a number,
       it is the time after the burst (typically listed for GCN only data).
       If letters, the first letter is for F = full light curve, I = interpolated between
          measured values on either side of 11 hr, and E = extrapolated from measured data. 
 The lower case letters indicate if the data in the referenced papers was in 
magnitude or Jansky units (mmf = first two references in magnitude units, 
third reference in Jansky units)}
\tablenotetext{d} {Optical data in $R-$band at 11 hr after the burst trigger. 
 $R-$band flux density estimated from typical burst spectra if data taken in other bands.  
Error includes 10\% systematic uncertainty.}

 
 \setcounter{refno}{0} 
\tablerefs{
\refstepcounter{refno} \label{tw1}     [\arabic{refno}] Cobb \& Bailyn (2005)
\refstepcounter{refno} \label{tw2}     [\arabic{refno}] Liang et al. (2008)
\refstepcounter{refno} \label{tw3a}     [\arabic{refno}] Wo\'zniak et al. (2005)
\refstepcounter{refno} \label{tw3b}     [\arabic{refno}] George et al. (2006)
\refstepcounter{refno} \label{tw3c}     [\arabic{refno}] Huang et al. (2007)
\refstepcounter{refno} \label{tw3d}     [\arabic{refno}] Kamble, Resmi, \& Misra (2007)
\refstepcounter{refno} \label{tw4a}     [\arabic{refno}] Ghirlanda, Nava, Ghisellini, \& Firmani (2007)
\refstepcounter{refno} \label{tw4b}     [\arabic{refno}] Soderberg  et al. (2007)
\refstepcounter{refno} \label{tw5a}     [\arabic{refno}] Shao \& Dai (2005)
\refstepcounter{refno} \label{tw5b}     [\arabic{refno}] Della Valle et al. (2006)
\refstepcounter{refno} \label{tw6}     [\arabic{refno}] Antonelli et al. (2006)
\refstepcounter{refno} \label{tw7a}     [\arabic{refno}] Pandey et al. (2006) 
\refstepcounter{refno} \label{tw7b}     [\arabic{refno}] Chen et al.  (2007)
\refstepcounter{refno} \label{tw8}     [\arabic{refno}] Rykoff et al. (2006)
\refstepcounter{refno} \label{tw10}     [\arabic{refno}] Cenko et al. (2006) 
\refstepcounter{refno} \label{tw11}     [\arabic{refno}] Sollerman et al. (2007)
\refstepcounter{refno} \label{tw13}     [\arabic{refno}] Durig \& Price (2005)
\refstepcounter{refno} \label{tw14}     [\arabic{refno}] Yost et al.  (2007)
\refstepcounter{refno} \label{tw16}     [\arabic{refno}] Oates et al. (2006)
\refstepcounter{refno} \label{tw17}     [\arabic{refno}] Misra et al. (2007) 
\refstepcounter{refno} \label{tw18a}     [\arabic{refno}] Monfardini et al. (2006)
\refstepcounter{refno} \label{tw18b}     [\arabic{refno}] Stanek et al. (2007)
\refstepcounter{refno} \label{tw18c}     [\arabic{refno}] Curran et al. (2007a)
\refstepcounter{refno} \label{tw19}     [\arabic{refno}] Curran et al. (2007b)
\refstepcounter{refno} \label{tw22}     [\arabic{refno}] Milne (2006)
\refstepcounter{refno} \label{tw23}     [\arabic{refno}] Dai et al. (2007)
\refstepcounter{refno} \label{tw24}     [\arabic{refno}] Garnavich \& Karska (2006)
\refstepcounter{refno} \label{tw26}     [\arabic{refno}] Nysewander et al. (2007)
\refstepcounter{refno} \label{tw27a}     [\arabic{refno}] Fynbo et al. (2006)
\refstepcounter{refno} \label{tw27b}     [\arabic{refno}] Mangano et al. (2007)
\refstepcounter{refno} \label{tw28}     [\arabic{refno}] Melandri, Tanvir \& Guidorzi (2006)
\refstepcounter{refno} \label{tw29}     [\arabic{refno}] Quimby \& Rykoff (2006)
\refstepcounter{refno} \label{tw30}     [\arabic{refno}] Soyano, Mito, \& Urata (2006)
\refstepcounter{refno} \label{tw31}     [\arabic{refno}] Mundell  et al. (2007)
\refstepcounter{refno} \label{tw32}     [\arabic{refno}] Thoene, Fynbo \& Jakobsson (2006)
\refstepcounter{refno} \label{tw33}     [\arabic{refno}] Fynbo  (2006)
\refstepcounter{refno} \label{tw34}     [\arabic{refno}] Page     et al. (2007)
\refstepcounter{refno} \label{tw35}     [\arabic{refno}] Perley   et al. (2007)
\refstepcounter{refno} \label{tw36}     [\arabic{refno}] Thoene   et al. (2007)
\refstepcounter{refno} \label{tw37}     [\arabic{refno}] Schmidt \& Mackie (2007)
\refstepcounter{refno} \label{tw38}     [\arabic{refno}] Thoene, Fynbo, \& Williams (2007)
\refstepcounter{refno} \label{tw39}     [\arabic{refno}] Terra    et al. (2007)
\refstepcounter{refno} \label{twx1}     [\arabic{refno}] Sharapov et al. (2005)
\refstepcounter{refno} \label{twx2}     [\arabic{refno}] Berger   et al. (2006)
\refstepcounter{refno} \label{twx3}     [\arabic{refno}] Efimov   et al. (2006)
\refstepcounter{refno} \label{twx6}     [\arabic{refno}] Malesani et al. (2007)
}
\end{deluxetable}

%% file: tab3.tex
\begin{deluxetable}{lccccccccc}
\tabletypesize{\tiny}
\tablewidth{0pt}
\tablecaption{{\it Swift} Long GRBs with X-ray Detections but No Optical
            Data at 11 Hr.}
\tablehead{
\colhead{(1)} & 
\colhead{(2)} & 
\colhead{(3)} & 
\colhead{(4)} & 
\colhead{(5)} & 
\colhead{(6)}  & 
\colhead{(7)}  & 
\colhead{(8)}  & 
\colhead{(9)}  & 
\colhead{(10)} \\ 
\colhead{GRB} &
\colhead{$\gamma$-ray\tablenotemark{a}} & 
\colhead{$\gamma$-ray\tablenotemark{a}} & 
\colhead{X-ray}  & 
\colhead{X-ray} &  
\colhead{X-ray} &  
\colhead{X-ray} &  
\colhead{X-ray} &  
\colhead{X-ray\tablenotemark{b}} &  
\colhead{X-ray\tablenotemark{b}} \\ 
\colhead{ }      &  
\colhead{Fluence} & 
\colhead{Fluence} & 
\colhead{Integral}& 
\colhead{Integral}& 
\colhead{Spectrum}& 
\colhead{Spectrum}& 
\colhead{Spectrum}& 
\colhead{Flux}  &   
\colhead{Flux}    \\ 
\colhead{ }  &  
\colhead{ } & 
\colhead{Error} & 
\colhead{Flux}& 
\colhead{Flux} & 
\colhead{Photon} & 
\colhead{Photon} & 
\colhead{Photon} & 
\colhead{Density}  & 
\colhead{Density} \\  
\colhead{ } & 
\colhead{ } & 
\colhead{ } & 
\colhead{ $0.3-10$ keV} & 
\colhead{Error} & 
\colhead{Index} & 
\colhead{Index} & 
\colhead{Index} & 
\colhead{@3 keV}  & 
\colhead{Error} \\ 
\colhead{ } & 
\colhead{ } & 
\colhead{ } & 
\colhead{ @11 hr} & 
\colhead{} & 
\colhead{} & 
\colhead{Lower} & 
\colhead{Upper} & 
\colhead{@11hr} & 
\colhead{} \\ 
\colhead{ } & 
\colhead{} & 
\colhead{} & 
\colhead{} & 
\colhead{} & 
\colhead{  } & 
\colhead{Error} & 
\colhead{Error} & 
\colhead{} & 
\colhead{} \\
\colhead{ } & 
\colhead{$10^{-7}$} & 
\colhead{$10^{-7}$} & 
\colhead{$10^{-13}$} & 
\colhead{$10^{-13}$} & 
\colhead{     } & 
\colhead{     } & 
\colhead{     } & 
\colhead{$10^{-3}$} & 
\colhead{$10^{-3}$} \\
\colhead{ } & 
\colhead{erg cm$^{-2}$} & 
\colhead{erg cm$^{-2}$} & 
\colhead{erg cm$^{-2}$ s$^{-1}$} & 
\colhead{erg cm$^{-2}$ s$^{-1}$} & 
\colhead{     } & 
\colhead{     } & 
\colhead{     } & 
\colhead{$\mu$Jy} & 
\colhead{$\mu$Jy} \\
}
\startdata
050124 & 11.9 & 0.66 & 12.4 & 2.7 & 1.89 & 0.22 & 0.27 & 51.3 & 14 \\
050128 & 50.2 & 2.3 & 23.6 & 5.2 & 2 & 0.19 & 0.21 & 92.8 & 26 \\
050215B & 2.27 & 0.29 & 2.76 & 1.1 & 1.67 & 0.4 & 0.47 & 12.3 & 5.3 \\
050219B & 158 & 5 & 38.8 & 3.6 & 2.01 & 0.15 & 0.16 & 151 & 22 \\
050223 & 6.36 & 0.65 & 1.28 & 0.53 & 1.9 & 0.51 & 0.62 & 5.26 & 2.7 \\
050326 & 88.6 & 1.6 & 12.1 & 2.5 & 2.05 & 0.21 & 0.45 & 46.1 & 15 \\
050505 & 24.9 & 1.8 & 31.1 & 3 & 2.03 & 0.085 & 0.089 & 120 & 15 \\
050603 & 63.6 & 2.3 & 27.6 & 3.6 & 1.93 & 0.11 & 0.12 & 113 & 18 \\
050607 & 5.92 & 0.55 & 2.13 & 0.65 & 2.49 & 0.5 & 0.59 & 5.67 & 4 \\
050712 & 10.8 & 1.2 & 11 & 2.2 & 2.18 & 0.23 & 0.26 & 38.4 & 12 \\
050713A & 51.1 & 2.1 & 22 & 2.9 & 2.27 & 0.15 & 0.17 & 71.8 & 15 \\
050713B & 31.8 & 3.2 & 77.5 & 12 & 1.89 & 0.17 & 0.19 & 321 & 65 \\
050714B & 5.95 & 1 & 2.63 & 0.72 & 2.88 & 0.38 & 0.21 & 4.43 & 2.7 \\
050716 & 61.7 & 2.4 & 9.57 & 1.7 & 2.16 & 0.25 & 0.29 & 33.9 & 10 \\
050726 & 19.4 & 2.1 & 4.38 & 0.88 & 2.11 & 0.25 & 0.29 & 16.1 & 4.9 \\
050814 & 20.1 & 2.2 & 10.2 & 1.7 & 2.01 & 0.14 & 0.15 & 40.1 & 8.6 \\
050819 & 3.5 & 0.55 & 2.75 & 1.1 & 2.44 & 0.46 & 0.57 & 7.72 & 5.6 \\
050822 & 24.6 & 1.7 & 17.9 & 2.3 & 2.21 & 0.15 & 0.16 & 61.1 & 12 \\
050915A & 8.5 & 0.88 & 3.63 & 0.89 & 1.93 & 0.33 & 0.43 & 14.8 & 5 \\
050915B & 33.8 & 1.4 & 5.74 & 1.9 & 2.21 & 0.3 & 0.37 & 19.7 & 9.3 \\
051001 & 17.4 & 1.5 & 2.3 & 0.51 & 2.46 & 0.28 & 0.35 & 6.31 & 2.8 \\
051008 & 50.9 & 1.4 & 9.78 & 1.6 & 2.16 & 0.19 & 0.2 & 34.6 & 8.6 \\
051016B & 1.7 & 0.22 & 11.9 & 2.4 & 1.86 & 0.19 & 0.21 & 49.7 & 13 \\
051117A & 43.4 & 1.6 & 5.96 & 1.1 & 2.36 & 0.18 & 0.21 & 18 & 5.3 \\
051221A & 11.5 & 0.35 & 7.52 & 1.4 & 2.12 & 0.17 & 0.19 & 27.5 & 6.9 \\
060109 & 6.55 & 1 & 4.96 & 1 & 2.58 & 0.3 & 0.35 & 12 & 5.7 \\
060111A & 12 & 0.58 & 5.1 & 0.78 & 2.32 & 0.18 & 0.21 & 15.9 & 4.2 \\
060111B & 16 & 1.4 & 3.73 & 0.9 & 2.14 & 0.29 & 0.35 & 13.5 & 5 \\
060115 & 17.1 & 1.5 & 5.66 & 2.3 & 2.72 & 0.46 & 0.68 & 11.6 & 10 \\
060202 & 21.3 & 1.6 & 10.6 & 1.3 & 3.21 & 0.17 & 0.19 & 11.2 & 3.9 \\
060204B & 29.5 & 1.8 & 7.18 & 1.3 & 2.33 & 0.22 & 0.26 & 22.3 & 7 \\
060211A & 15.7 & 1.4 & 2.84 & 0.79 & 2.47 & 0.34 & 0.43 & 7.73 & 4.2 \\
060306 & 21.3 & 1.2 & 11.7 & 1.5 & 2.28 & 0.15 & 0.17 & 37.8 & 8.1 \\
060319 & 2.64 & 0.34 & 8.65 & 1.4 & 2.21 & 0.18 & 0.22 & 29.5 & 7.6 \\
060428A & 13.9 & 0.78 & 66 & 8 & 2.21 & 0.19 & 0.2 & 226 & 49 \\
060428B & 8.23 & 0.81 & 3.22 & 0.54 & 1.92 & 0.16 & 0.18 & 13.2 & 2.8 \\
060507 & 44.5 & 2.3 & 7.85 & 1.6 & 2.14 & 0.21 & 0.25 & 28.4 & 8.3 \\
060510A & 80.5 & 3.1 & 178 & 21 & 1.98 & 0.069 & 0.14 & 708 & 100 \\
060510B & 40.7 & 1.8 & 1.45 & 0.5 & 2.32 & 0.36 & 0.5 & 4.54 & 2.6 \\
060707 & 16 & 1.5 & 15.4 & 4.1 & 2.05 & 0.25 & 0.33 & 59 & 21 \\
060708 & 4.94 & 0.37 & 7.5 & 1 & 2.05 & 0.12 & 0.12 & 28.6 & 5 \\
060712 & 12.4 & 2.2 & 3.09 & 0.74 & 2.45 & 0.25 & 0.32 & 8.59 & 3.6 \\
060719 & 15 & 0.91 & 5.93 & 1.2 & 2.77 & 0.27 & 0.33 & 11.5 & 5.6 \\
060804 & 5.98 & 0.99 & 15.1 & 3.3 & 2.26 & 0.25 & 0.35 & 49.7 & 18 \\
060807 & 8.48 & 1.1 & 9.29 & 1.3 & 2.43 & 0.19 & 0.21 & 26.2 & 7.3 \\
060813 & 54.6 & 1.4 & 43.2 & 11 & 2.16 & 0.33 & 0.36 & 154 & 61 \\
060814 & 146 & 2.4 & 31.4 & 3.2 & 2.21 & 0.11 & 0.12 & 107 & 16 \\
060923A & 8.69 & 1.3 & 5.44 & 1 & 2.07 & 0.19 & 0.28 & 20.5 & 5.6 \\
060923C & 15.8 & 2.2 & 4.05 & 1 & 2.72 & 0.4 & 0.5 & 8.34 & 5.7 \\
061004 & 5.66 & 0.31 & 2.06 & 0.67 & 2.4 & 0.53 & 0.67 & 6.04 & 4.2 \\
061019 & 25.9 & 4 & 12.8 & 3 & 2.05 & 0.38 & 0.45 & 49 & 19 \\
061222A & 79.9 & 1.6 & 158 & 13 & 2.09 & 0.095 & 0.099 & 587 & 68 \\
070103 & 3.38 & 0.46 & 0.951 & 0.21 & 2.06 & 0.27 & 0.29 & 3.62 & 1.1 \\
070107 & 51.7 & 2.6 & 58 & 5.7 & 2.2 & 0.14 & 0.14 & 200 & 33 \\
070129 & 29.8 & 2.7 & 16.5 & 4.6 & 2.14 & 0.22 & 0.25 & 59.5 & 22 \\
070208 & 4.45 & 1 & 2.92 & 0.89 & 2.46 & 0.43 & 0.54 & 8.01 & 5.1 \\
070220 & 104 & 2.3 & 7.26 & 1.5 & 1.7 & 0.23 & 0.27 & 32.2 & 7.7 \\
070223 & 17 & 1.2 & 4.02 & 1.2 & 1.9 & 0.65 & 0.95 & 16.6 & 7.9 \\
070306 & 53.8 & 2.9 & 77.8 & 11 & 2.18 & 0.16 & 0.18 & 273 & 60 \\
070318 & 24.8 & 1.1 & 12.8 & 1.9 & 2.33 & 0.17 & 0.19 & 40 & 9.9 \\
070328 & 90.6 & 1.8 & 45.6 & 5.5 & 2.03 & 0.14 & 0.16 & 176 & 30 \\
070330 & 1.83 & 0.31 & 4.62 & 1.2 & 2.37 & 0.41 & 0.54 & 13.9 & 7.7 \\
070420 & 140 & 4.5 & 53.9 & 5.2 & 2.04 & 0.16 & 0.17 & 207 & 32 \\
070521 & 80.1 & 1.8 & 21.1 & 2.8 & 1.98 & 0.18 & 0.2 & 84 & 16 \\
070529 & 25.7 & 2.4 & 3.73 & 0.87 & 2.18 & 0.26 & 0.4 & 13.1 & 4.9 \\
070611 & 3.91 & 0.57 & 2.06 & 0.59 & 1.95 & 0.29 & 0.35 & 8.31 & 3.1 \\
070616 & 192 & 3.5 & 8.41 & 1.6 & 2.49 & 0.24 & 0.3 & 22.3 & 8.7 \\
070621 & 43. & 1. & 7.85 & 1.4 & 2.63 & 0.29 & 0.33 & 17.8 & 8.2 \\
070704 & 59. & 3. & 6.69 & 1.8 & 1.97 & 0.35 & 0.58 & 26.7 & 11 \\
070714A & 1.5 & 0.2 & 1.53 & 0.53 & 2.42 & 0.73 & 0.95 & 4.36 & 3.8 \\
070721A & 0.71 & 0.18 & 1.43 & 0.45 & 2.65 & 0.39 & 0.39 & 3.17 & 2.1 \\
070721B & 36. & 2. & 3.5 & 0.72 & 1.88 & 0.18 & 0.17 & 14.6 & 3.7 \\
\enddata
\tablenotetext{a}{BAT prompt fluence in $15-150$ keV band. Data from Sakamoto et al. (2008).}
\tablenotetext{b}{XRT flux density at 3 keV at 11 hr after the burst trigger. Error includes 10\% systematic uncertainty.}

 
\end{deluxetable}

%% file: tab4.tex
\begin{deluxetable}{lcccccc}
\tabletypesize{\tiny}
\tablewidth{0pt}
\tablecaption{Correlation Fits and Coefficients}
\tablehead{
\colhead{(1)} & 
\colhead{(2)} & 
\colhead{(3)} & 
\colhead{(4)} & 
\colhead{(5)} & 
\colhead{(6)}  & 
\colhead{(7)} \\ 
\colhead{Data Set}  & 
\colhead{Number of} & 
\colhead{A\tablenotemark{a}} & 
\colhead{B\tablenotemark{a}}  & 
\colhead{Correlation} &  
\colhead{Null} &        
\colhead{Fraction of} \\ 
\colhead{ }      &  
\colhead{Data} & 
\colhead{ } & 
\colhead{ }& 
\colhead{Coefficient}& 
\colhead{Hypothesis} &  
\colhead{variability} \\ 
\colhead{ }  &  
\colhead{Points } & 
\colhead{ } & 
\colhead{ }& 
\colhead{ } & 
\colhead{Probability\tablenotemark{b}} & 
\colhead{due to} \\   
\colhead{ } & 
\colhead{ } & 
\colhead{ } & 
\colhead{ } & 
\colhead{ } & 
\colhead{ } & 
\colhead{correlation\tablenotemark{b}} \\  
\colhead{ } & 
\colhead{$N$} & 
\colhead{ } & 
\colhead{ } & 
\colhead{$r$} & 
\colhead{$P_{\rm null}$} & 
\colhead{$r^2$ } \\ 
}
\startdata
   Long GRBs:        &      &                &                 &      &           &       \\
  Optical  ($y$)  vs X-ray ($x$)    & 37  & $1.62\pm0.04$ & $0.38\pm0.03$  & $0.44\pm0.03$  & 0.006  & 0.19  \\
                     &      &                &                 &      &              &       \\
   Short GRBs:       &      &                &                 &      &               &       \\
  Optical  ($y$)   vs X-ray ($x$)   &  6  & $0.72\pm0.94$ & $0.14\pm0.45$ & $0.06\pm0.23$ & 0.68 & 0.00  \\
                     &      &                &                 &      &                 &       \\
   Long GRBs:        &      &                &                 &      &                 &       \\
 X-ray    ($y$) vs $\gamma$-ray ($x$)& 111 & $2.11\pm0.21$ & $0.63\pm0.04$ & $0.53\pm0.02$ & $4\times10^{-9}$& 0.28  \\
                     &      &                &                 &      &                 &       \\
  Short GRBs:        &      &                &                 &      &                 &       \\
 X-ray    ($y$)  vs $\gamma$-ray ($x$)& 10 & $0.06\pm1.07$ & $0.36\pm0.17$ & $0.35\pm0.14$     &  0.31      & 0.12  \\
                     &      &               &                 &      &                 &       \\
  Long  GRBs:        &      &               &                 &      &                 &       \\
  Fluence   ($y$) vs Peak Flux ($x$) & 218 & $-6.03\pm0.01$ & $0.83\pm0.02$ & $0.66\pm0.01$ & $4\times10^{-29}$& 0.44  \\
                     &      &               &                 &      &                 &       \\
 Short  GRBs:        &      &               &                 &      &                 &       \\
 Fluence   ($y$)  vs Peak Flux  ($x$) & 17 & $-7.06\pm0.04$ & $1.27\pm0.06$ & $0.84\pm0.02$ & $2\times10^{-6}$  & 0.71  \\
                     &      &              &                 &      &                 &       \\
\enddata
\tablenotetext{a}{Fit with the function $y=10^A x^B$}
\tablenotetext{b}{The significance of the correlation is $1-P_{\rm null}$.}

 
\end{deluxetable}

%% file: ms.bbl
\begin{thebibliography}{75}
\expandafter\ifx\csname natexlab\endcsname\relax\def\natexlab#1{#1}\fi

\bibitem[]{}
Antonelli, L.~A., et al. 2006, A\&A, 456, 509

\bibitem[]{}
Barthelmy, S.~D., et al. 2005a, Space Sci Rev, 120, 143

\bibitem[]{}
Barthelmy, S.~D., et al. 2005b, Nature, 438, 994

\bibitem[]{}
Berger, E.,  et al. 2005, Nature, 438, 988

\bibitem[]{}
Berger, E.,  et al. 2006, GCN Circ. 5697, http://gcn.gsfc.nasa.gov/gcn/gcn3/5697.gcn3


\bibitem[]{}
Berger, E. 2007, ApJ, 670, 1254

\bibitem[]{}
Berger, E., et al. 2007, ApJ, 664, 1000

\bibitem[]{}
Bikmaev, I., et al. 2005, GCN Circ. 3797, http://gcn.gsfc.nasa.gov/gcn/gcn3/3797.gcn3

\bibitem[]{}
Bloom, J.~S., et al. 1999, Nature, 401, 453

\bibitem[]{}
Bloom, J.~S., et al. 2006, ApJ, 638, 354

\bibitem[]{}
Burrows, D.~N., et al. 2005, Space Sci Rev, 120, 165

\bibitem[]{}
Butler, N.~R. 2007, ApJ, 656, 1001



\bibitem[]{}
Cenko, S.~B., et al. 2006, ApJ, 652, 490

\bibitem[]{}
Chen, H.-W., et al. 2007, ApJ, 663, 420

\bibitem[]{}
Cobb, B.~E., \& Bailyn, C.~D. 
                   2005, GCN Circ. 3104, http://gcn.gsfc.nasa.gov/gcn/gcn3/3104.gcn3

\bibitem[]{}
Curran, P.~A., et al. 2007a, MNRAS, 381, L65

\bibitem[]{}
Curran, P.~A., et al. 2007b, A\&A, 467, 1049



\bibitem[]{}
Dai, X., et al. 2007, ApJ, 658, 509

\bibitem[]{}
D'Avanzo, P., et al.
                   2006, GCN Circ. 5884, http://gcn.gsfc.nasa.gov/gcn/gcn3/5884.gcn3

\bibitem[]{}
Della Valle, M., et al. 2006, ApJ, 642, L103

\bibitem[]{}
de Pasquale, M.,  et al.
          2003, ApJ, 592, 1018

\bibitem[]{}
de Pasquale, M., et al.
          2006, A\&A, 455, 813

\bibitem[]{}
Durig, D.~T., \& Price, A.
                   2005, GCN Circ. 4023, http://gcn.gsfc.nasa.gov/gcn/gcn3/4023.gcn3

\bibitem[]{}
Efimov, Yu., et al. 2006, GCN Circ. 5986, http://gcn.gsfc.nasa.gov/gcn/gcn3/5986.gcn3

\bibitem[]{}
Eichler, D., Livio, M., Piran, T., \& Schramm, D.~N. 1989, Nature, 340, 126

\bibitem[]{}
Fenimore, E., et al. 2004, AIP, 727, 667

\bibitem[]{}
Firmani, C., Ghisellini, G., Avila-Reese, V., 
          \& Ghirlanda, G. 2006,
                MNRAS, 370, 185

\bibitem[]{}
Fox, D.~B., et al. 2005, Nature, 437, 845

\bibitem[]{}
Fukugita, M., Shimasaku, K., \&  Ichikawa, T. 1995, PASP, 107, 945

\bibitem[]{}
Fynbo, J.~P.~U., et al. 2007, astro-ph/0703458

\bibitem[]{}
Fynbo, J.~P.~U., et al. 2006, Nature, 444, 1047

\bibitem[]{}
Galama, T.~J., et al. 1998, Nature, 395, 670

\bibitem[]{}
Garnavich, P., \& Karska, A. 2006,
                        GCN Circ. 5253, http://gcn.gsfc.nasa.gov/gcn/gcn3/5253.gcn3

\bibitem[]{}
Gehrels, N., et al. 2004, ApJ, 611, 1005

\bibitem[]{}
Gehrels, N., et al. 2005, Nature, 437, 859

\bibitem[]{}
Genet, F., Daigne, F., \& Mochkovitch, R. 2007, MNRAS, 381, 732

\bibitem[]{}
George, K., Banerjee, D.~P.~K., Chandrasekhar, T., \& Ashok, N.~M. 2006, ApJ, 640, L13

\bibitem[]{}
Ghirlanda, G., Nava, L., Ghisellini, G., \& Firmani, C. 2007, A\&A, 466, 127

\bibitem[]{}
Ghisellini, G., Ghirlanda, G., Nava, L., \& Firmani, C. 2007, ApJ, 658, L75

\bibitem[]{}
Grupe, D. et al. 2007, ApJ, 662, 443



\bibitem[]{}
Hjorth, J., et al. 2003, Nature, 423, 847

\bibitem[]{}
Hjorth, J., et al. 2005, Nature, 437, 851

\bibitem[]{}
Huang, K.~Y., et al. 2007, ApJ, 654, L25

\bibitem[]{}
Jakobsson, P., et al. 2004, ApJ, 617, L21

\bibitem[]{}
Kamble, A., Resmi, L., \& Misra, K. 2007, ApJ, 664, L5

\bibitem[]{}
Kouveliotou, C., Meegan, C.~A., Fishman, G.~J., Bhat, N.~P., Briggs, M.~S., Koshut, T.~M.,
     Paciesas, W.~S., \& Pendleton, G.~N. 1993, ApJ, 413, L101

\bibitem[]{}
Kulkarni, S.~R. 2005, astro-ph/0510256

\bibitem[]{}
Lattimer, J.~M., \& Schramm, D.~N. 1974, ApJ, 192, L145

\bibitem[]{}
Lee, W.~H., \& Ramirez-Ruiz, E. 2007, New J. Phys., 9, 17


\bibitem[]{}
Li, L.-X., \& Paczy\'nski, B. 1998, ApJ, 507, L59

\bibitem[]{}
Liang, E.-W., Racusin, J.~L., Zhang, B., Zhang, B.-B., 
           \& Burrows, D.~N.  
              2008, ApJ, 675, 528

\bibitem[]{}
Liang, E.-W., Zhang, B.-B., \& Zhang, B. 2007, ApJ, 670, 565

\bibitem[]{}
MacFadyen, A.~I., \& Woosley, S.~E. 1999, ApJ, 524, 262

\bibitem[]{}
Malesani, D., Stella, L., Covino, S., Lidman, C., \& Naef, D. 2006,
                           GCN Circ. 5705, http://gcn.gsfc.nasa.gov/gcn/gcn3/5705.gcn3

\bibitem[]{}
Malesani, D., et al. 2007, GCN Circ. 6565, http://gcn.gsfc.nasa.gov/gcn/gcn3/6565.gcn3

\bibitem[]{}
Malesani, D., et al. 2007, GCN Circ. 6674, http://gcn.gsfc.nasa.gov/gcn/gcn3/6674.gcn3

\bibitem[]{}
Malesani, D., et al. 2007, A\&A, 473, 77

\bibitem[]{}
Mangano, V. et al. 2007, A\&A, 470, 105


\bibitem[]{}
Melandri, A., Tanvir, N., \& Guidorzi, C.
                         2006, GCN Circ. 5322, http://gcn.gsfc.nasa.gov/gcn/gcn3/5322.gcn3

\bibitem[]{}
M\'esz\'aros, P.,  \& Rees, M.~J. 1997, ApJ, 476, 232

\bibitem[]{}
Milne, P.~A.                2006, GCN Circ. 5127, http://gcn.gsfc.nasa.gov/gcn/gcn3/5127.gcn3

\bibitem[]{}
Misra, K., \& Pandey, S.~B. 2005, GCN Circ. 3396, http://gcn.gsfc.nasa.gov/gcn/gcn3/3396.gcn3

\bibitem[]{}
Misra, K., et al. 2007, A\&A, 464, 903

\bibitem[]{}
Mochkovitch, R., Hernanz, M., Isern, J., \& Martin, X. 1993, Nature 361, 236

\bibitem[]{}
Monfardini, A., et al. 2006, ApJ, 648, 1125

\bibitem[]{}
Mundell, C.~G., et al.
          2006, GCN Circ. 5700,
          http://gcn.gsfc.nasa.gov/gcn/gcn3/5700.gcn3


\bibitem[]{}
Mundell, C.~G., et al. 2007, ApJ, 660, 489


\bibitem[]{}
Nakar, E., et al. 2007, Phys Reports, 442, 166

\bibitem[]{}
Nava, L., Ghisellini, G., Ghirlanda, G., Tavecchio, F., 
       \& Firmani, C.
          2006, A\&A, 450, 471

\bibitem[]{}
Nousek, J.~A., et al. 2006, ApJ, 642, 389

\bibitem[]{}
Nysewander, M., et al.                     2007, astro-ph/0708.3444v2

\bibitem[]{}
Nysewander, M., Fruchter, A.~S., \& Pe'er, A. 2008, astro-ph/0806.3607v1

\bibitem[]{}
Oates, S.~R., et al. 2006, MNRAS, 372, 327

\bibitem[]{}
Oechslin, R., Janka, H.-T., \& Marek, A. 2007, A\&A, 467, 395

\bibitem[]{}
Paczy\'nski, B. 1986, ApJ, 308, L43

\bibitem[]{}
Page, K.~L., et al. 2007, ApJ, 663, 1125

\bibitem[]{}
Panaitescu, A. 2008, MNRAS, 383, 1143

\bibitem[]{}
Panaitescu, A., et al. 2006, MNRAS, 366, 1357

\bibitem[]{}
Pandey, S.~B., et al. 2006, A\&A, 460, 415

\bibitem[]{}
Perley, D.~A., et al. 2008, ApJ, 672, 449

\bibitem[]{}
Perley, D.~A., Thoene, C. C., \& Bloom, J. S. 2007, 
                   GCN Circ. 6774, http://gcn.gsfc.nasa.gov/gcn/gcn3/6774.gcn3

\bibitem[]{}
Pian, E.,  et al.  2006, Nature, 442, 1011

\bibitem[]{}
Piran, T., Kumar, P., Panaitescu, A., \& Piro, L. 2001, ApJ, 560, L167

\bibitem[]{}
Press, W.~H., Flannery, B.~P., Teukolsky, S.~A., \& Vetterling, W.~T.
      1986, Numerical Recipes (New York: Cambridge Univ. Press)

\bibitem[]{}
Price, P.~A., et al.         2006, GCN Circ. 5077, http://gcn.gsfc.nasa.gov/gcn/gcn3/5077.gcn3

\bibitem[]{}
Quimby, R., \& Rykoff, E. S. 2006, GCN Circ. 5377, http://gcn.gsfc.nasa.gov/gcn/gcn3/5377.gcn3

\bibitem[]{}
Rol, E., et al. 2005, ApJ, 624, 868

\bibitem[]{}
Rol, E., et al. 2007, ApJ, 669, 1098

\bibitem[]{}
Roming, P.~W.~A., et al. 2005, Space Sci Rev, 120, 95

\bibitem[]{}
Roming, P.~W.~A., et al. 2006, ApJ, 652, 1416

\bibitem[]{}
Rosswog, S., Ramirez-Ruiz, E., \& Davies, M.~B. 2003, MNRAS, 345, 1077

\bibitem[]{}
Rykoff, E.~S., et al. 2006, ApJ, 638, L5

\bibitem[]{}
Sakamoto, T., et al. 2006, in ``Gamma-Ray Bursts in the {\it Swift} Era'', 
ed. S.~S. Holt, N. Gehrels, and J.~A. Nousek (AIP: New York), p. 43

\bibitem[]{}
Sakamoto, T., et al. 2008, ApJS, 175, 179

\bibitem[]{}
Salmonson, J.~D., \& Galama, T.~J. 2002, ApJ, 569, 682

\bibitem[]{}
Schlegel, D.~J., Finkbeiner, D.~P., \& Davis, M. 1998, ApJ, 500, 525

\bibitem[]{}
Schmidt, B., \& Bayliss, D. 2006, 
                                 GCN Circ. 4880, http://gcn.gsfc.nasa.gov/gcn/gcn3/4880.gcn3

\bibitem[]{}
Schmidt, B., \& Mackie, G. 2007, GCN Circ. 6325, http://gcn.gsfc.nasa.gov/gcn/gcn3/6325.gcn3

\bibitem[]{}
Shao, L., \& Dai, Z.~G. 2007, ApJ, 660, 1319

\bibitem[]{}
Sharapov, D. et al. 2005,        GCN Circ. 3701, http://gcn.gsfc.nasa.gov/gcn/gcn3/3701.gcn3

\bibitem[]{}
Shao, L., \& Dai, Z.~G. 2005, ApJ, 633, 1027

\bibitem[]{}
Soderberg, A.~M., et al. 2006, ApJ, 650, 261

\bibitem[]{}
Soderberg, A.~M., et al. 2007, ApJ, 661, 982

\bibitem[]{}
Sollerman, J., et al. 2007, A\&A, 466, 839

\bibitem[]{}
Soyano, T., Mito, H., \& Urata, Y. 2006, GCN Circ. 5548,
http://gcn.gsfc.nasa.gov/gcn/gcn3/5548.gcn3

\bibitem[]{}
Spearman, C. 1904, Am J Psychol, 15, 72

\bibitem[]{}
Stanek, K.~Z., et al. 2003, ApJ, 591, L17

\bibitem[]{}
Stanek, K.~Z., et al. 2007, ApJ, 654, L21

\bibitem[]{}
Stefanescu, A., et al. 2006, GCN Circ. 5291, http://gcn.gsfc.nasa.gov/gcn/gcn3/5291.gcn3

\bibitem[]{}
Stratta, G., Fiore, F., Antonelli, L.~A., Piro, L., 
       \& de Pasquale, M. 2004, ApJ, 608, 846

\bibitem[]{}
Terra, F., et al. 2007, GCN Circ., 6458, http://gcn.gsfc.nasa.gov/gcn/gcn3/6458.gcn3

\bibitem[]{}
Thoene, C.~C., Fynbo, J.~P.~U., \& Jakobsson, P. 2006, GCN Circ. 5747,
http://gcn.gsfc.nasa.gov/gcn/gcn3/5747.gcn3

\bibitem[]{}
Thoene, C.~C., Fynbo, J.~P.~U., \& Williams, A. 2007, GCN Circ. 6389,
http://gcn.gsfc.nasa.gov/gcn/gcn3/6389.gcn3

\bibitem[]{}
Thoene, C.~C., Kann, D.~A., Augusteijn, T., \& Reyle-Laffont, C.
2007, GCN Circ. 6154, http://gcn.gsfc.nasa.gov/gcn/gcn3/6154.gcn3

\bibitem[]{}
Uhm, Z.~L., \& Beloborodov, A.~M. 2007, ApJ, 665, L93

\bibitem[]{}
van der Horst, A.~J., et al. 2008, ApJ, in prep.


\bibitem[]{}
Villasenor, J.~S., et al. 2005, Nature, 437, 851.

\bibitem[]{}
Willingale, R., et al. 2007, ApJ, 662, 1093

\bibitem[]{}
Woosley, S.~E., 1993, ApJ, 405, 273.

\bibitem[]{}
Woosley, S.~E., \& Bloom, J.~S. 2006, ARAA, 44, 507

\bibitem[]{}
Wo\'zniak, P.~R., Vestrand, W.~T., Wren, J.~A., 
White, R.~R., Evans, S.~M., \& Casperson, D. 2005, ApJ, 627, L13

\bibitem[]{}
Yost, S.~A., et al. 2007, ApJ, 657, 925

\bibitem[]{}
Zhang, B., et al. 2006, ApJ, 642, 354

\bibitem[]{}
Zhang, B., et al. 2007, ApJ, 655, 989

\bibitem[]{}
Zhang, Z.-B., \& Choi, C.-S. 2008, A\&A, 484, 293

\bibitem[]{}
Zhang, B., \& M\'esz\'aros, P. 2004, IJMPA, 19, 2385

\end{thebibliography}
